\documentclass[trackchanges,twocolumn]{aastex701}

\usepackage[version=4]{mhchem}
\usepackage{siunitx} 
\usepackage[export]{adjustbox}
\newcommand{\tint}{$T_{\rm int}$} 
\newcommand{\teq}{$T_{\rm eq}$}
\newcommand{\tp}{$T(P)$}

\newcommand{\co}{CO}
\newcommand{\meth}{CH$_4$}

\newcommand{\amon}{NH$_3$}
\newcommand{\cotwo}{CO$_2$}

\newcommand{\RNum}[1]{\uppercase\expandafter{\romannumeral #1\relax}}
\begin{document}

%\title{On the Effects of Clouds in Sub-Neptune Atmospheres and Interiors}
\title{Impact of Clouds on the Atmosphere-Mantle Interface of Sub-Neptunes}

\author[0000-0003-1622-1302]{Sagnick Mukherjee}\email[show]{smukhe50@asu.edu}
\altaffiliation{51 Pegasi b Fellow} 
\affiliation{School of Earth and Space Exploration, Arizona State University, Tempe, AZ, USA \\ }

\author[0000-0001-8236-5553]{Matthew C.\ Nixon}\email[]{}
\altaffiliation{51 Pegasi b Fellow} 
\affiliation{School of Earth and Space Exploration, Arizona State University, Tempe, AZ, USA \\ }

\author[0000-0003-1622-1302]{Luis Welbanks}\email[]{}
\affiliation{School of Earth and Space Exploration, Arizona State University, Tempe, AZ, USA \\ }

\author[0000-0001-5864-9599]{James Mang}\email[]{}
\altaffiliation{NSF Graduate Research Fellow.}
\affiliation{Department of Astronomy, University of Texas at Austin, Austin, TX, USA\\}

\author[0000-0002-0413-3308]{Nicholas F. Wogan}\email[]{}
\affiliation{NASA Ames Research Center, Moffett Field, CA 94035\\}
\affiliation{SETI Institute, Mountain View, CA, USA\\}

\author[0000-0003-1240-6844]{Natasha E. Batalha}\email[]{}
\affiliation{NASA Ames Research Center, Moffett Field, CA, USA\\}

\author[0000-0001-6247-8323]{Michael R. Line}\email[]{}
\affiliation{School of Earth and Space Exploration, Arizona State University, Tempe, AZ, USA \\ }

\begin{abstract}

Sub-Neptunes are among the most common type of close-in planets found in our galaxy, yet their bulk composition remains largely uncertain;  H-rich envelopes overlaying rocky cores, volatile-rich planets, and carbon-rich interiors all remain viable configurations for members of this population. Atmospheric characterization has been proposed as a means of distinguishing between these scenarios, but growing evidence suggests that sub-Neptunes may host molten atmosphere-mantle interfaces which could alter the composition of their atmosphere. We use the \texttt{PICASO} 1D climate model, coupled to interior-structure and magma-atmosphere chemistry frameworks to quantify how clouds alter the atmospheric and interior structure of sub-Neptunes. For temperate sub-Neptunes like TOI-270~d, we find that clouds can lead to $\ge{1000}$~K heating at depth (${\sim}10^{4}$~bar) and $\sim{600}$~K cooling at shallow pressures ($\sim$1~bar). This heating is very sensitive to the cloud sedimentation efficiency and, to a lesser extent, to metallicity. Most sub-Neptunes in our sample should have a molten atmosphere-mantle interface, except TOI-1231~b and GJ~1214~b. For these two planets, cloudy models have a molten interface whereas clear models can allow a solid boundary. Clouds can heat the atmosphere-mantle interfaces by a temperature difference between $\sim{1400}-2600$~K for most sub-Neptunes in our sample. Such cloud-driven heating can substantially change the composition of the interface with abundances of \ce{O2}, \ce{SiH4}, and \ce{SiO} showing a $\ge{36}$\% increase between cloudy and clear models of TOI-270~d. We discuss the implications of our results for the thermal evolution and measurements of intrinsic heat flux for this population.

\end{abstract}

\keywords{\uat{Exoplanets}{498}, \uat{Exoplanet atmospheres}{487}}%\uat{Solar physics}{1476}}


\section{Introduction} 

% Sub-Neptune exoplanets are among the most common outcomes of the complex planet formation and evolution process in our galaxy \citep{fulton2017, howard,fressin13}. Their radii lie between \numrange{1.7}{4}~$R_{\oplus}$ \citep{rogers2015}. Radial velocity campaigns of their host stars and measurements of transit timing variations in these planetary systems have been employed to measure their masses  \citep[e.g.,][]{luque2022a,Turtelboom22,Rosenthal21,jiang2025,livingstone26}. These measurements have revealed that sub-Neptunes have densities that can be explained by multiple compositions and interior structures \citep[e.g.,][]{adams2008,luque2022a}.

% A possibility is that these planets have a voluminous H-rich envelope, comprising only a small fraction of their mass, overlaying a rocky core \citep[e.g.,][]{barnes2009,rogers2010,rogers2015,lopez14}. On the other hand, ``water-worlds" composed of $\ge$10\% \ce{H2O} by mass \citep[e.g.,][]{Kuchner2003,Fu2010,fortney2007planetary,Rogers2010b,Zeng2019} or even carbon-rich compositions \citep{Lin2025,Li2026,madhu12} are also consistent with their measured mass and radii. Atmospheric characterization has been proposed to be the key to distinguish between these scenarios \citep[e.g.,][]{adams2008,millerricci2009,fortney13}.

 Sub-Neptunes are the most common close-in planets in the galaxy \citep{fulton2017, howard,fressin13}. Their radii lie between \numrange{1.7}{4}~$R_{\oplus}$ \citep{rogers2015}, and their measured bulk densities \citep[e.g.,][]{luque2022a,Turtelboom22,Rosenthal21,jiang2025,livingstone26} are consistent with bulk structures as varied as H-rich envelopes overlaying rocky cores \citep[e.g.,][]{barnes2009,rogers2010,rogers2015,lopez14}, `water-worlds" composed of $\ge$10\% \ce{H2O} by mass \citep[e.g.,][]{Kuchner2003,Fu2010,fortney2007planetary,Rogers2010b,Zeng2019}, and carbon-rich compositions \citep{Lin2025,Li2026,madhu12}. Mass and radius alone cannot distinguish these scenarios, and atmospheric characterization has been proposed as a means to distinguish them \citep[e.g.,][]{adams2008,millerricci2009,fortney13}.

The strength of gaseous absorption features in transmission spectra of sub-Neptunes are a function of their atmospheric mean molecular weight (along with temperature, gravity, and aerosol-driven obscuration), which might be indicative of the chemical composition of these planets \citep{benneke2013}. For example, a very \ce{H2O}-rich composition might have a much larger atmospheric mean molecular weight compared to a planet with an extended H/He envelope over a rocky or ice-rich core. However, physical processes in the atmospheres and their interaction with planetary interiors might further complicate this picture.

Atmospheric processes including condensation \citep{morley13,charnay21,gao18,lee2025}, photochemical production of hazes \citep{He24,mak2025,gao2017sulfur}, photochemistry \citep{tsai23,mukherjee25}, and large-scale dynamics \citep{Mukherjee22a,zhang18a,fortney20} can significantly alter the chemical composition, as well as the mean molecular weight, of planetary photospheres. Apart from these atmospheric processes, deep interiors of sub-Neptunes might develop molten magma oceans due to extremely high pressures and temperatures at the base of their envelopes \citep{kite2019,kite2020,tang25,Nixon2025}. The possibility of water oceans has also been proposed for temperate sub-Neptunes \citep{madhu2021,Nixon2021,piette2020}.

Magma oceans can exchange gases at the mantle-envelope interface, which can alter their atmospheric chemical composition \citep{hakim26,Misener2023,schlichting2022,kite2020,horn2025,Nixon2025} and modify their thermal structures \citep{Misener2023,markham2022,misener2022}. Understanding the physical nature of this atmosphere-mantle interface \citep{breza25,calder2025,tejada2026} is therefore critical for leveraging atmospheric observations of sub-Neptunes to reveal their composition, evolution, and formation mechanisms.

Observational efforts to characterize sub-Neptune atmospheres with space-based telescopes have been ongoing for more than a decade \citep[e.g.,][]{Kreidberg2014,brande24,benneke2019,barat24,mikalevans2021,mikalevans2023}. Aerosol-driven muting of their transmission spectra  emerged as a common theme across these observations, albeit at varying levels \citep{dymont2022,brande24}. Recently, {\it JWST} has enabled transmission spectroscopy of these sub-Neptune atmospheres across a much wider wavelength range with enhanced precision \citep[e.g.,][]{madhu23,hu2025,benneke2024,davenport25,gao23,barat2025,piaulet2024,wallack24,wallack2026,Teske25}. Featureless spectra or muted spectral features have remained the most prominent results borne out of these studies, which has been postulated as due to either high mean molecular weight atmospheres or atmospheres dominated by aerosols. Phase curve observations of sub-Neptunes have further highlighted the role of aerosols  in their atmospheres \citep{kempton23}. 

% While a significant volume of theoretical work has sought to understand the nature of these aerosols \citep[e.g.,][]{he2024,horst2018,Huseby2025}, their composition \citep[e.g.,][]{morley13,gao18,lavvas19}, and spatial distribution \citep[e.g.,][]{steinrueck2025,malsky2025}, how aerosol coverage in sub-Neptunes can impact the atmosphere-mantle interface has remained underexplored. \citet{lee2025} explored mineral cloud formation in sub-Neptune exoplanets, finding abundant cloud formation near the atmosphere-mantle boundary. They also suggest that these clouds might have substantial impact on the boundary itself but these effects were not quantified. The goal of this study is to theoretically quantify how such mineral clouds lurking deep beneath the observable photosphere (for transmission spectroscopy) can alter the physical conditions at the atmosphere-mantle interface for a wide variety of sub-Neptune exoplanets.\

While a significant volume of theoretical work has sought to understand the nature of these aerosols \citep[e.g.,][]{he2024,horst2018,Huseby2025}, their composition \citep[e.g.,][]{morley13,gao18,lavvas19}, and spatial distribution \citep[e.g.,][]{steinrueck2025,malsky2025}, how aerosol coverage in sub-Neptunes can impact the atmosphere-mantle interface has remained underexplored. \citet{lee2025} explored mineral cloud formation in sub-Neptune exoplanets, finding abundant cloud formation near the atmosphere-mantle boundary and suggested these clouds could substantially impact the boundary, but did not quantify the effect. Here, we quantify the impact of mineral clouds on the pressure, temperature, phase, and chemistry of the atmosphere-mantle interface across a representative sample of sub-Neptune exoplanets.

We describe our atmosphere and interior structure modeling in \S\ref{sec:modeling}, highlight our results in \S\ref{sec:results}, discuss the implications of our results in \S\ref{sec:disc}, and present our conclusions in \S\ref{sec:conc}.

\section{Modeling}\label{sec:modeling}

%\subsection{Atmosphere Modeling}
\subsection{Climate and Chemistry}
We use \texttt{PICASO} \citep{Mukherjee22,mang26,batalha19} to simulate sub-Neptune atmospheres. Our climate model assumes radiative-convective equilibrium and is self-consistently coupled to the {\it photochem} 1D chemical kinetics model \citep{wogan2024,wogan2025,mukherjee25}. We divide our model atmospheres into 90 plane-parallel layers with pressures equally-spaced in the logarithmic space. We assume a full-redistribution of the incident stellar energy between the day- and night- sides of the model planets, which corresponds to \texttt{rfacv}=0.5 in \texttt{PICASO}. We assume $T_{\rm int}=50$ K for all models, in line with the relatively cold interiors expected from the mature sub-Neptune population due to their thermal evolution \citep{fortney20,fortney2007planetary,thorngren2016,tang25}. 

We compute atmospheric chemistry by interfacing the radiative-convective model ``on-the-fly" with the {\it photochem} 1D chemical kinetics model. {\it Photochem} computes the abundances of gases at each major iteration of the {\tp} profile within \texttt{PICASO}. We use a C-H-N-O-S chemical network to compute the atmospheric chemistry for \ce{CO}, \ce{CH4}, \ce{H2O}, \ce{NH3}, \ce{CO2}, \ce{N2}, \ce{HCN}, \ce{H2}, \ce{C2H2}, \ce{C2H4}, \ce{C2H6}, \ce{SO2}, and \ce{H2S} with {\it photochem}. On the other hand, we compute the abundances of \ce{Na}, \ce{K}, and \ce{FeH} assuming thermochemical equilibrium with the equilibrium chemistry solver within {\it photochem}. 

We use mixing length theory \citep{gierasch85} to compute the strength of vertical mixing, parametrized using the 1D eddy diffusion coefficient-- $K_{\rm zz}$, in the convective atmospheric layers. In the radiative layers, the $K_{\rm zz}$ is also computed using mixing length theory but the mixing length is decreased using a factor, which is the ratio of the local lapse rate and adiabatic gradient following previous work \citep{diamondback,ackerman2001cloud}. This $K_{\rm zz}$ profile is used to compute the chemical state in disequilibrium with {\it photochem} at each iteration of the climate model.

Correlated-k opacities of \ce{CO}, \ce{CH4}, \ce{H2O}, \ce{NH3}, \ce{CO2}, \ce{N2}, \ce{HCN}, \ce{H2}, \ce{C2H2}, \ce{C2H4}, \ce{C2H6}, \ce{SO2}, \ce{H2S}, \ce{Na}, \ce{K}, and \ce{FeH} are mixed at every iteration of the climate model with the resort-rebin technique \citep{amundsen17}. We do not include photolysis reactions in {\it photochem}, and instead only consider gas-phase kinetics because the effects of photochemical processes on the $T(P)$ profile are known to be small ($\sim$100 K) and localized to smaller pressures \citep{mukherjee25}. Therefore, the {\it photochem} model computes the abundance profiles taking quenching due to vertical mixing and thermochemical equilibrium into account.
\subsection{Clouds}
We simulate clouds using the \texttt{VIRGA} cloud model \citep{batalha25,rooney21}. The \texttt{VIRGA} cloud model derives its heritage from the \texttt{eddysed} cloud model \citep{ackerman2001cloud} and has been fully coupled with the climate modules of \texttt{PICASO 4.0} \citep{mang26}. This coupling ensures that the cloud structure and its impact on the thermal structure are computed self-consistently within our climate model, enabling us to explore radiatively active mineral clouds in sub-Neptune atmospheres.

\texttt{VIRGA} simulates the vertical distribution of cloud droplets by solving the balance between the vertical lofting of cloud droplets and condensible vapor due to atmospheric dynamics and the gravitational settling of droplets. This balance is described by,

\begin{equation}
    -K_{\rm zz}\dfrac{\partial{q_{\rm t}}}{\partial{z}}-f_{\rm sed}w_*q_c=0
\end{equation}
where $q_{\rm t}$ is the mass mixing ratio of the cloud droplets and condensible vapor, $z$ is altitude, $f_{\rm sed}$ is the ratio between the settling velocity of droplets and the convective velocity in the atmosphere ($w_*$), and $q_{\rm c}$ is the mass mixing ratio of the cloud droplets \citep{ackerman2001cloud}. Lower $f_{\rm sed}$ causes cloud droplets to be extended to lower pressures or higher altitudes, whereas a higher $K_{\rm zz}$ causes bigger cloud droplets to remain lofted in the atmosphere. We do not vary $K_{\rm zz}$ as a free parameter but calculate it self-consistently to preserve the physical self-consistency between the climate and cloud modeling framework \citep{diamondback}. \texttt{VIRGA} assumes a log-normal distribution of droplet sizes. The droplet size distribution at each atmosphere layer is set by the  mean particle radii and its standard deviation. These are calculated using both $K_{\rm zz}$ and $f_{\rm sed
}$ following \citet{ackerman2001cloud}. We use the same $K_{\rm zz}$ profile for cloud droplets and vertical transport of gases.

We consider seven cloud species \ce{MgSiO3}, \ce{Fe}, \ce{Al2O3}, \ce{Na2S}, \ce{KCl}, \ce{MnS}, and \ce{ZnS} that have been predicted to condense in the deeper atmospheres of warm to temperate exoplanets \citep{morley2012neglected,visscher06,visscher10,lee2025}. We use wavelength-dependent cloud optical properties of each of these species (Appendix \ref{sec:thermalappen}) and assume perfectly spherical cloud droplets. We also consider a spatially uniform coverage of clouds, even though there is growing empirical evidence that cloud coverage in giant exoplanets can be patchy \citep{Mukherjee2025,Fu2025,coulombe2025,MacDonald2017,Welbanks2021}. We further discuss this assumption in \S\ref{sec:disc}.

%\subsection{Atmosphere-Interior Interface Modeling}
\subsection{Interior Structure Modeling} \label{subsec:interior}

We model the internal structure of sub-Neptunes using \texttt{SMILE} \citep{Nixon2021, Nixon2024}, which computes the radius of a planet for a given mass, $T(P)$ profile, and composition. \texttt{SMILE} is used to solve the equations of hydrostatic equilibrium and mass conservation for planets consisting of an iron core, silicate mantle, and mixed gaseous envelope consisting of H, He and H$_2$O. The atmospheric $T(P)$ profile and composition are taken from the \texttt{PICASO} models. Due to limited equation of state data for many of the chemical species in the atmosphere, we use H$_2$O as a proxy for all atmospheric components heavier than H/He, a common assumption in many internal structure models \citep[e.g.,][]{thorngren2016,Egger2025,Rigby2026}. We choose the proportions of atmospheric H/He and H$_2$O such that the mean molecular weight profile is identical to that of the atmospheric models, following \citet{Nixon2024}. In cases where the gaseous envelope extends to deeper pressures than considered in the atmospheric model, we assume an adiabatic $T(P)$ profile and a constant mean molecular weight, fixed to its value at the deepest pressure in the atmospheric model.

For each planet under consideration, we explore a grid of models with varying envelope mass fractions from 0.1\% to 50\% in steps of 0.05\%. We use steps of 0.005\% for TOI-421~b due to its smaller bulk envelope fraction compared to all other targets. We accept all models for which the model mass and radius are consistent with measured values to 1$\sigma$. For each accepted model, we extract the pressure and temperature at the atmosphere-mantle boundary, to find the range of surface conditions that may be possible for the planet.

\subsection{Atmosphere-Interior Exchange}

We also investigate how changes in pressures and temperatures at the atmosphere-mantle interface due to deep atmospheric clouds can impact the chemical products of magma-atmosphere interactions. For the majority of targets considered in this work, the surface of the silicate layer is expected to be molten (magma), meaning that chemical interactions between the silicate layer and the overlying gaseous envelope can significantly alter the planet's chemical composition \citep{Schaefer2016,kite2019}. The resulting composition at the base of the atmosphere is therefore strongly dependent on the pressure-temperature conditions at the interface \citep{schlichting2022,Nixon2025}.

We use the equilibrium thermodynamic model for a sub-Neptune-like exoplanet which accounts for chemical interactions between a metal core, silicate-dominated mantle, and an initially hydrogen-rich atmosphere presented in \citet{schlichting2022}. The temperature at the atmosphere-mantle boundary is chosen to match the values extracted from the internal structure models, as described in Section \ref{subsec:interior}. We employ the “reactive metal” model, in which the metal (possibly in the form of a differentiated core) participates fully in the chemistry. The system is described by a basis set of reactions, and the condition for chemical equilibrium is a zero sum of the product of the chemical potentials of each species and their stoichiometric coefficients in each reaction.

\section{Results}\label{sec:results}
% First, we explore climate models for TOI-270~d to examine how clouds affect the {\tp} structure relative to clear models, and how this effect depends on the sedimentation efficiency of cloud droplets and on atmospheric metallicity (\S\ref{sec:toi270d}). {\bf We then explore the impact of clouds on the deep atmosphere over a generalized gravity--$T_{\rm eq}$ grid before focusing on a sample of 8 representative sub-Neptunes, including TOI-270~d, where we explore how mineral clouds alter the phase of atmosphere-mantle interfaces, their $T$-$P$ conditions, and their chemical composition (\S\ref{sec:interface} \& \S\ref{sec:chemical})}.

We first explore how clouds affect the temperature-pressure structure of the temperate sub-Neptune TOI-270~d  (\S\ref{sec:toi270d}), and how this effect depends on the sedimentation efficiency of cloud droplets and atmospheric metallicity. Then, we explore the impact of cloud-driven heating over a generalized gravity--$T_{\rm eq}$ grid to later focus on a sample of 8 representative sub-Neptunes, including TOI-270~d, for which we quantify the impact of mineral clouds on their atmosphere-mantle interfaces their phases and pressure-temperature conditions (\S\ref{sec:interface}), and their chemical composition (\S\ref{sec:chemical})).

\subsection{Radiative feedback of clouds in sub-Neptune atmospheres}\label{sec:toi270d}
\begin{figure*}
    \centering
    \includegraphics[width=1\linewidth]{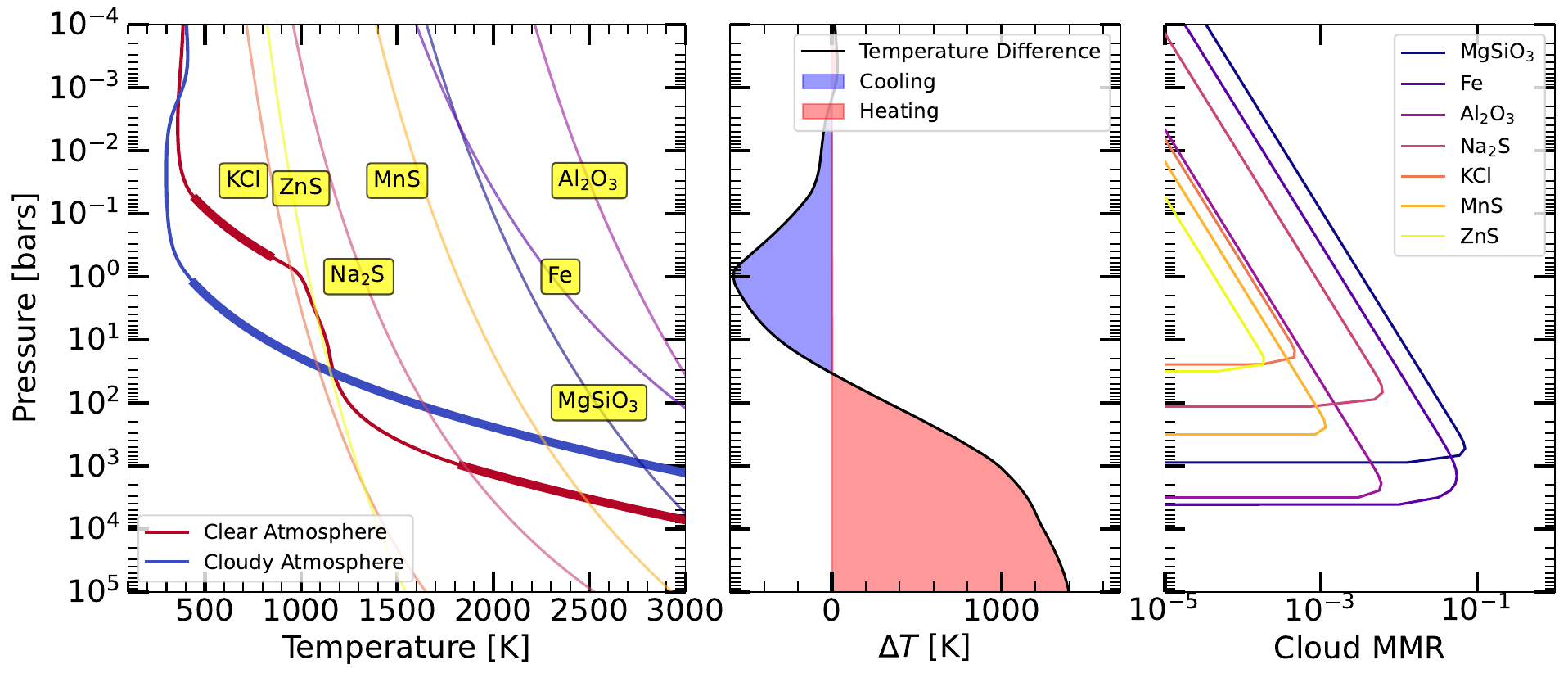}
    \caption{{\it Left panel}: The {\tp} profiles for a clear (red line) and cloudy (blue line) model for TOI-270~d's atmosphere are shown along with condensation curves for several mineral cloud species. The thicker regions depict convective regions of the atmosphere and the models have been computed for [M/H]=+2.3 and {$f_{\rm sed}=0.5$}. {\it Middle panel}: The temperature difference between the cloudy and clear atmospheres as a function of pressure. {\it Right panel}: Mass mixing ratio profiles as a function of pressure for the mineral cloud species considered in our models.}
    \label{fig:cloudvsclear}
\end{figure*}

% We simulate cloudy and clear self-consistent atmosphere models for the temperate sub-Neptune TOI-270~d \citep{Gunther2019}. We assume a significantly metal enriched atmosphere with [M/H]$=+2.3$ and a C/O$=0.23$, which are consistent with the derived composition from {\it JWST} observations of the planet \citep{benneke2024}. We assume $f_{\rm sed}=0.5$ for the cloudy model, but also explore the effects of higher or lower sedimentation efficiencies in \S\ref{sec:fsed}. 

 We use TOI-270~d \citep{Gunther2019} as our primary case study because its atmospheric metallicity and C/O ratio have recently been constrained by {\it JWST} observations \citep{benneke24}, allowing for our cloudy and clear self-consistent models to be anchored against these empirical estimates. We assume a significantly metal enriched atmosphere with [M/H]$=+2.3$ and a C/O$=0.23$, consistent with the derived composition from those observations \citep{benneke2024}. We assume $f_{\rm sed}=0.5$ for the cloudy model, but also explore the effects of higher or lower sedimentation efficiencies in \S\ref{sec:fsed}.

Figure~\ref{fig:cloudvsclear} shows the difference in the atmospheric structure of TOI-270~d between a cloudy and a clear atmosphere model. The left panel of Figure~\ref{fig:cloudvsclear} compares the {\tp} profiles between the two cases. The condensation curves of the considered cloud species are also shown. The cloudy model has a colder {\tp} profile between pressures \numrange{0.001}{30}~bar but a much hotter {\tp} profile for $P\ge{30}$~bar. We also find that the clear atmosphere model shows a deep radiative-convective boundary near $\sim{1000}$~bar pressure, whereas the radiative-convective boundary for the cloudy model extends to $\sim{1}$~bar, which is three orders of magnitude shallower in pressure. However, the clear atmosphere model shows a detached convective region near $\sim$0.1~bar. This difference in the thermal structure between these two cases is driven by two effects: clouds reflecting incident stellar light to space, and clouds trapping the thermal radiation in the planet's deep atmosphere. We explore the effect of these two processes in Appendix \ref{sec:thermalappen}.

The temperature difference ($\Delta{T}$) between these two models is shown in the middle panel of Figure~\ref{fig:cloudvsclear}. We find that clouds can heat up the atmosphere of TOI-270~d by about 1200~K near $10^{4}$~bar and cool their shallower atmospheres (near $1$~bar) by $\sim$600~K. This suggests that deep mineral clouds can cause additional condensation at higher altitudes of temperate sub-Neptune atmospheres. For example, this cooling effect near smaller pressures could lead to \ce{H2O} condensation in planets for which clear-atmosphere models do not predict it. We further discuss the impact of these clouds on the transmission spectra of temperate sub-Neptune TOI-270~d in \S\ref{disc:spec}.  The right panel of Figure~\ref{fig:cloudvsclear} shows the mass mixing ratio profiles of the condensates. The maximum mass mixing ratios of condensates like \ce{Fe} or \ce{MgSiO3} can reach as high as $\sim$10\% at pressures near ${\sim}1000$~bar due to elevated metallicities ($200\times$solar here).

\subsubsection{Role of cloud droplet settling efficiency}\label{sec:fsed}

\begin{figure*}
    \centering
    \includegraphics[width=1\linewidth]{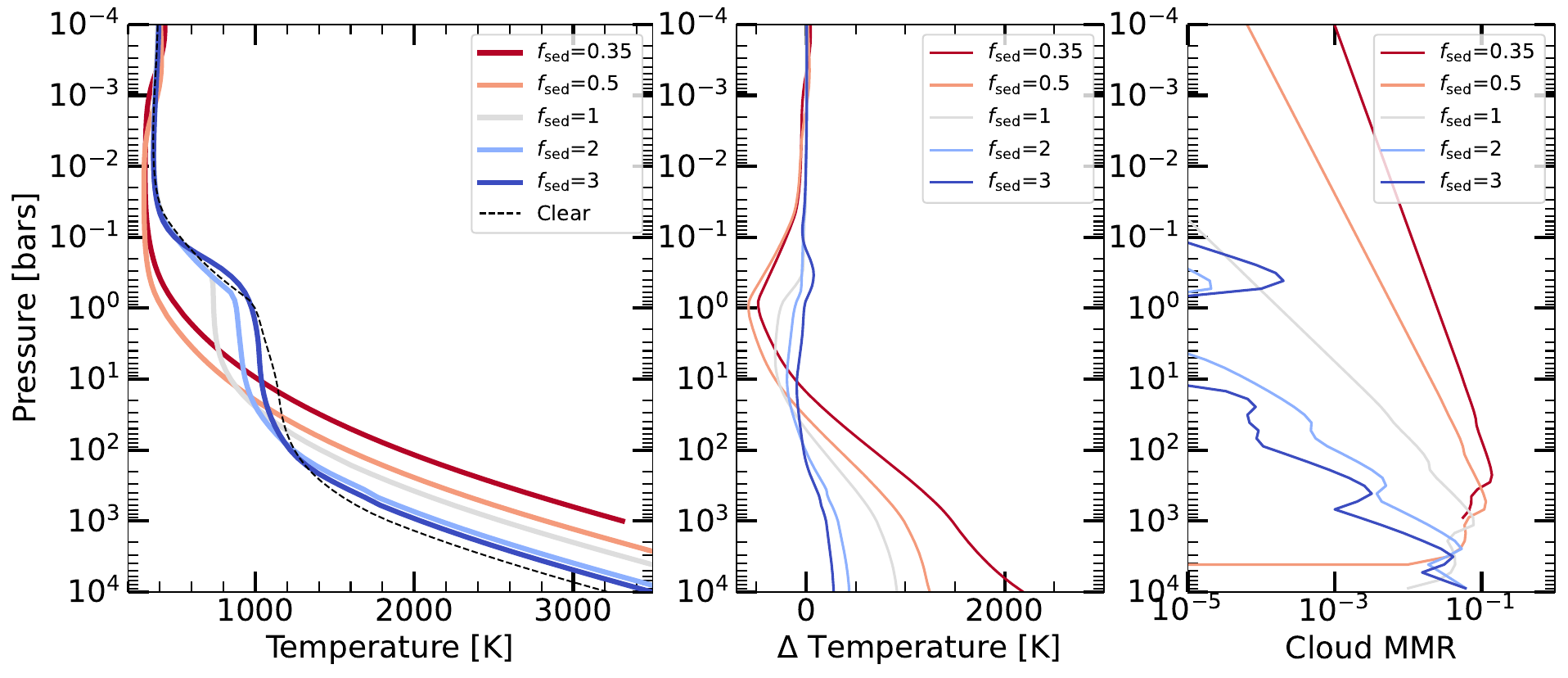}
    \caption{{\it Left panel}: {\tp} profiles of TOI-270~d for various cloud sedimentation parameter $f_{\rm sed}$ values between 3 to 0.35 are shown with the solid colored lines. The {\tp} profile for the clear atmosphere model is shown with the dashed black line. {\it Middle panel}: Difference in temperature between the cloudy atmosphere model and the clear atmosphere model for each $f_{\rm sed}$ value is shown as colored lines. A positive value suggests that the cloudy model is hotter than the clear model at that pressure. {\it Right panel}: Variation of the total cloud mass mixing ratio with $f_{\rm sed}$ is shown. The mass mixing ratios of all individual gases are summed together to get the total cloud mass mixing ratio.}
    \label{fig:fsed}
\end{figure*}

% The settling rate of cloud droplets has remained unconstrained in exoplanet atmospheres \citep{diamondback,rooney21,Gao+18_BD}. We explore the effect of this settling rate on the {\tp} profile of TOI-270~d. The settling rate in the \texttt{VIRGA} cloud model is parametrized with $f_{\rm sed}$ (\S\ref{sec:modeling}). We explore five different $f_{\rm sed}$ values between 0.35 and 3, representing typical values used for Solar System planets and exoplanets \citep{Gao+18_BD}. The left panel of Figure~\ref{fig:fsed} compares the five different {\tp} profiles from these cloudy models with a clear model for TOI-270~d. The middle panel shows the change in temperature in these cloudy atmospheres relative to the clear model, while the right panel shows the mass mixing ratio profiles of clouds. The mass mixing ratio profiles show that the cloud droplets extend to lower pressures for lower $f_{\rm sed}$ values. As $f_{\rm sed}$ decreases, the temperatures at smaller pressures become progressively colder relative to the clear model. On the other hand, the peak mass mixing ratio of clouds increases with decreasing $f_{\rm sed}$. This causes the deep cloudy atmospheres to get increasingly hot relative to the clear atmosphere with decreasing $f_{\rm sed}$. 

The settling rate of cloud droplets, parametrized in \texttt{VIRGA} as $f_{\rm sed}$ (\S\ref{sec:modeling}), determines the vertical extent of the clouds and the depth to which their radiative effects propagate. This rate is poorly constrained in exoplanet atmospheres \citep{diamondback,rooney21,Gao+18_BD}. We explore five different $f_{\rm sed}$ values between 0.35 and 3, representing typical values used for Solar System planets and exoplanets \citep{Gao+18_BD}. The left panel of Figure~\ref{fig:fsed} compares the five different {\tp} profiles from these cloudy models with a clear model for TOI-270~d. The middle panel shows the change in temperature in these cloudy atmospheres relative to the clear model, while the right panel shows the mass mixing ratio profiles of clouds. The mass mixing ratio profiles show that the cloud droplets extend to lower pressures for lower $f_{\rm sed}$ values. As $f_{\rm sed}$ decreases, the temperatures at smaller pressures become progressively colder relative to the clear model. On the other hand, the peak mass mixing ratio of clouds increases with decreasing $f_{\rm sed}$. This causes the deep cloudy atmospheres to get increasingly hot relative to the clear atmosphere with decreasing $f_{\rm sed}$.

The cloudy atmosphere with $f_{\rm sed}=0.35$ is hotter by $\sim{2000}$~K relative to the clear atmosphere at $10{^4}$~bar, whereas this heating only causes $\sim{250}$~K temperature difference for $f_{\rm sed}=3$. This suggests that if cloud particles settle very efficiently, then they should have little impact on the deep atmospheric structure and the atmosphere-mantle boundary. However, if the settling velocity of cloud droplets are low relative to the strength of vertical dynamics, they could have a dramatic effect on the deep atmospheric conditions and the atmosphere-mantle interface. 

 Recent observations of transiting exoplanets, directly imaged exoplanets, and brown dwarfs show that cloud droplets are lofted to low to very low pressures  \citep{Mukherjee2025,welbanks24,kiefer2026,suarez22,Mukherjee2025gj436b,gao23}. The $f_{\rm sed}$ measured from fitting the {\it JWST} spectra of several exoplanets have found values that are often smaller than 1 \citep[e.g.,][]{inglis24,Mukherjee2025,kiefer2026,grant23}. Additionally, sub-Neptune transmission spectra show strongly muted spectral features \citep{gao23,brande24,wallack24,wallack2026} and their phase-curve measurements have revealed high albedos as well \citep{kempton23}. This suggests that the expected $f_{\rm sed}$ values in these atmospheres could be low rather than high. Therefore, we assume $f_{\rm sed}=0.5$ for the rest of this work as a nominal case.  

\subsubsection{Variation of cloud radiative feedback with  metallicity}

\begin{figure*}
    \centering
    \includegraphics[width=1\linewidth]{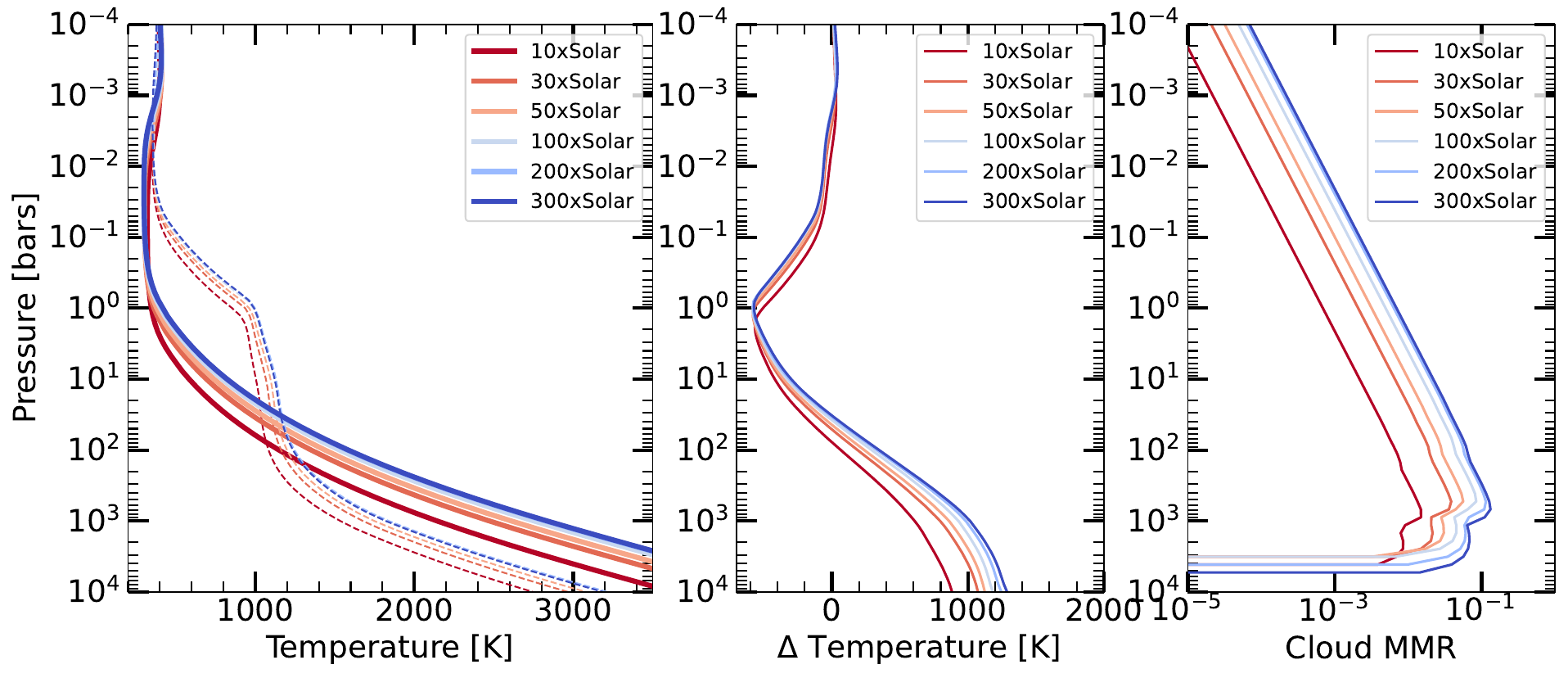}
    \caption{{\it Left panel}: {\tp} profiles of TOI-270~d for various atmospheric metallicities between 10$\times$solar and 300$\times$solar are shown as colored lines. The {\tp} profile for the clear atmosphere model for each metallicity are shown with the dotted lines and those for the cloudy models are shown with solid lines. All models are computed assuming $f_{\rm sed}=0.5$. {\it Middle panel}: Difference in temperature between the cloudy atmosphere model and the clear atmosphere model for each metallicity value. A positive value suggests that the cloudy model is hotter than the clear model at that pressure. {\it Right panel}: Variation of the total cloud mass mixing ratio with metallicity. The mass mixing ratios of all individual gases are summed together to get the total cloud mass mixing ratio.}
    \label{fig:metal}
\end{figure*}

% The mineral cloud species we have considered are composed of elements other than H and He (e.g., \ce{MgSiO3}, \ce{Na2S}, \ce{ZnS}, \ce{Fe}). As the atmospheric metallicity increases, the abundance of condensable species will also increase. Figure~\ref{fig:metal} shows how clouds impact the thermal structure of TOI-270~d relative to clear models at various atmospheric metallicities (for $f_{\rm sed}=0.5$). The left panel presents the clear {\tp} profiles (dashed lines) and the cloudy {\tp} profiles (solid lines) at various metallicities between 10$\times$ to 300$\times$solar. The middle panel shows the temperature differences between the cloudy and clear models at each metallicity and the right panel shows the cloud mass mixing ratio profiles. 

The cloud species we have considered are composed of elements other than H and He (e.g., \ce{MgSiO3}, \ce{Na2S}, \ce{ZnS}, \ce{Fe}), so the abundance of condensable species (and therefore the cloud reservoir) increases with atmospheric metallicity. Figure~\ref{fig:metal} shows how clouds impact the thermal structure of TOI-270~d relative to clear models at various atmospheric metallicities (for $f_{\rm sed}=0.5$). The left panel presents the clear {\tp} profiles (dashed lines) and the cloudy {\tp} profiles (solid lines) at various metallicities between 10$\times$ to 300$\times$solar. The middle panel shows the temperature differences between the cloudy and clear models at each metallicity and the right panel shows the cloud mass mixing ratio profiles. 

The maximum cloud mass mixing ratio increases, as expected, with increasing metallicity from $\sim$1\% at 10$\times$solar to $\ge$10\% at 300$\times$solar. As a result, the heating caused by clouds in the deep atmosphere ($\sim{10^4}$~bar) increases with increasing metallicity. The temperature difference caused by clouds increases from about ${900}$~K at 10$\times$solar metallicity to about ${1300}$~K at 300$\times$solar metallicity at $10^{4}$~bar owing to increasing cloud mass mixing ratio with increasing metallicity. The right panel in Figure~\ref{fig:metal} shows that the maximum mass mixing ratio of clouds shift from higher pressures to smaller pressures with increasing metallicity, which is also driven by the hotter deep atmospheres in higher metallicity cases. Therefore, this heating impact of clouds is likely much more important in the deep atmospheres of metal-rich planets (e.g., sub-Neptunes) compared to gas giants that are nominally expected to be less rich in metals \citep{welbanks19}.

\subsubsection{Impact throughout the ${T_{\rm eq}}$ vs.\ $log(g)$ space}

\begin{figure}
    \centering
    \includegraphics[
        width=1\linewidth
    ]{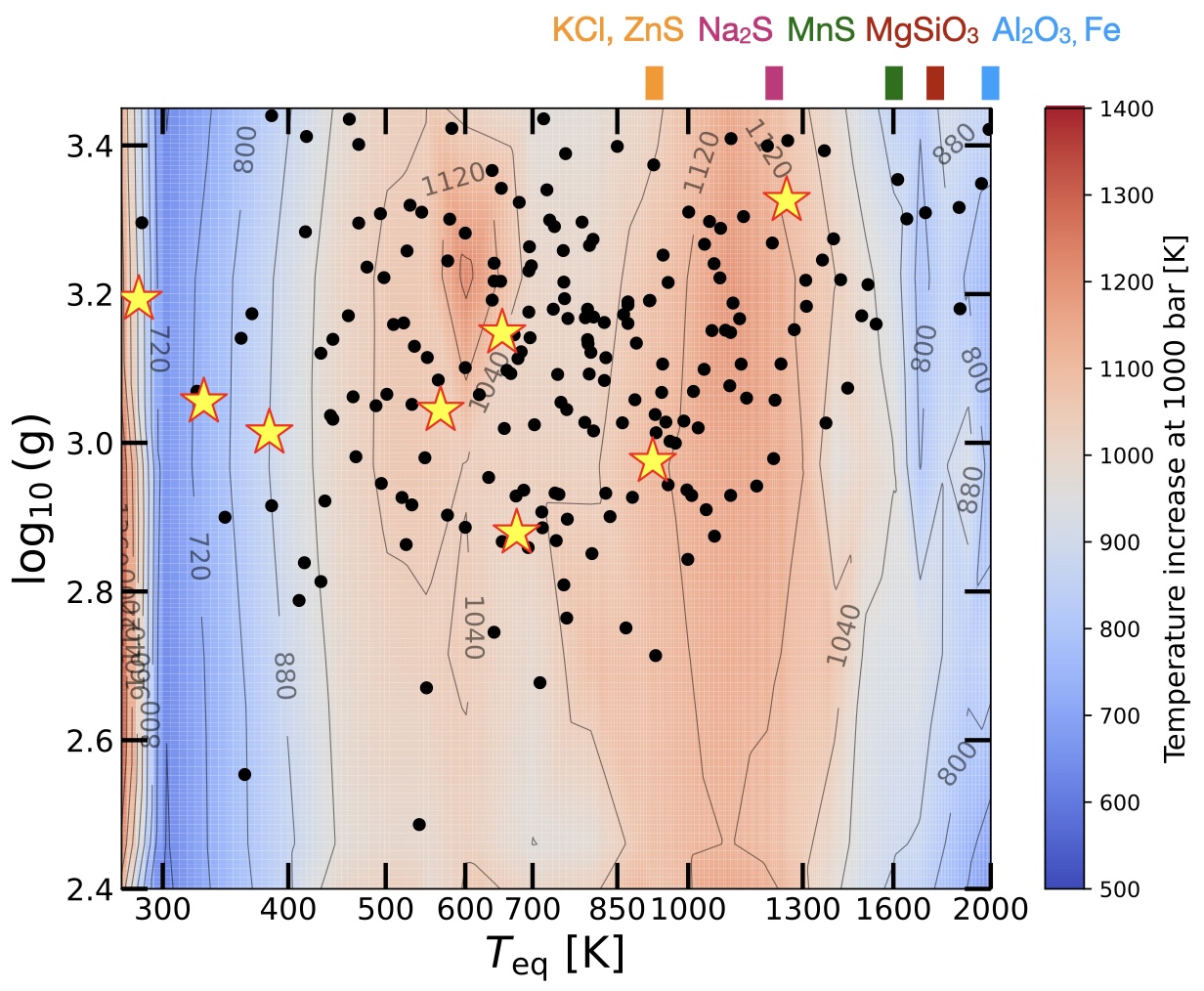}
    \caption{The difference in temperature at 1000~bar pressure between cloudy and clear models is shown as a function of planet $T_{\rm eq}$ and $log(g)$. The cloudy and clear grid of models have been computed assuming 100$\times$Solar metallicity, solar C/O ratio, and $T_{\rm int}=50$~K after adopting the stellar properties of TOI-270~d. The cloudy models assume $f_{\rm sed}=0.5$. The $T_{\rm eq}$ at which each cloud species starts forming clouds for $log(g)=3.0$ have been marked at the top of the panel with colored vertical lines. Black points show the sample of transiting sub-Neptune planets with 1.7${\le}R_{\rm p}\le$4 R$_{\oplus}$. Planets that have radius and mass measurements with $\le{25}$\% uncertainty are shown here. The yellow stars show our sample of 8 sub-Neptunes, which span the range of $T_{\rm eq}$ and $\log(g)$ occupied by the broader sub-Neptune population and for which we have explored the impact of clouds on the atmosphere-mantle interface. }
    \label{fig:sample}
\end{figure}
%\begin{figure}
%    \centering
%    \includegraphics[width=1\linewidth]{figures/sample_grid.pdf}
%    \caption{The difference in temperature at 1000~bar pressure between cloudy and clear models is shown as a function of planet $T_{\rm eq}$ and $log(g)$. The cloudy and clear grid of models have been computed assuming 100$\times$Solar metallicity, solar C/O ratio, and adapting the stellar properties of TOI-270~d. The cloudy models assume $f_{\rm sed}=0.5$. Black points show the sample of transiting sub-Neptune planets with 1.7${\ge}R_{\rm p}\ge$4 R$_{\oplus}$. Planets that have radius and mass measurements with $\le{25}$\% uncertainty are shown here. The yellow stars show our sample of 8 sub-Neptunes, which span the range of $T_{\rm eq}$ and $\log(g)$ occupied by the broader sub-Neptune population and for which we explored the impact of clouds on the atmosphere-mantle interface. }
%    \label{fig:sample}
%\end{figure}

 The above case study shows that clouds can profoundly impact the temperature at depth in sub-Neptune atmospheres. Here, we generalize this result by first exploring how the heating effects of clouds vary with planetary $T_{\rm eq}$ and gravity, followed by planet-specific analyses for representative, well studied planets. We generate a grid of clear models with $T_{\rm int}=50$~K, varying $T_{\rm eq}$ from 250~K to 2000~K and planet gravity from 1.6~ms$^{-2}$ to 30~ms$^{-2}$. We adopt the stellar properties of TOI-270 and assume an atmospheric metallicity of 100{$\times$}Solar. We also compute a grid of cloudy models within this parameter space, assuming $f_{\rm sed}=0.5$. The heat map in Figure~\ref{fig:sample} shows the temperature difference between the cloudy and clear models as a function of $T_{\rm eq}$ and $log(g)$ at 1000 bar. The cloudy models are always $\ge700$~K hotter than the clear models at this pressure.

 The nearly vertical contours in Figure~\ref{fig:sample} show that this heating is sensitive to $T_{\rm eq}$ but not to the planet's gravity. We find that lowering gravity makes both the clear and cloudy models hotter at deep pressures (e.g., 1000 bar), but by nearly the same amount. The dependance of the heating on $T_{\rm eq}$ is shaped by the onset of condensation of various clouds with lowering temperatures and subsequent sinking of those clouds in the atmosphere. The cloud species that dominates the heating at a given $T_{\rm eq}$ depends on $log(g)$, as changing gravity shifts the $T(P)$ profile relative to the condensation curves. However, beyond $T_{\rm eq}{\sim}1800$~K, the heating is generally driven by Al$_2$O$_3$ and Fe clouds. The temperature difference caused by clouds increases from about 800~K to $\sim$1100~K when $T_{\rm eq}$ decreases from 2000~K to 1200~K due to the onset of MgSiO$_3$ and MnS condensation. This level of heating is maintained by Na$_2$S, ZnS, and KCl clouds for 1200~K$\ge{T_{\rm eq}}\ge$400~K. For $T_{\rm eq}\le$400~K, the cloud-driven temperature difference at 1000~bar decreases to $\sim$700~K owing to the sinking of these cloud decks to deeper pressures with colder temperatures.

 The sample of transiting sub-Neptunes (1.7${\le}R_{\rm p}\le$4 R$_{\oplus}$) that have mass and radius uncertainties smaller than 25\% are shown as black points in Figure~\ref{fig:sample}. The impact of clouds at 1000 bar is substantial across the parameter space occupied by the sub-Neptune population. To explore how clouds impact the interior-mantle interface in sub-Neptunes, we pick a sample of 8 representative planets spanning this $T_{\rm eq}$ vs.\ $log(g)$ space. We investigate the impact of clouds on the atmosphere-mantle interface for-- K2-18~b \citep{montet2015}, TOI-270~d \citep{Gunther19}, TOI-421~b \citep{carleo2020}, TOI-824~b \citep{burt2020}, TOI-836~c \citep{hawthorn2023}, TOI-1231~b \citep{Burt2021}, GJ~9827~d \citep{Niraula2017}, and GJ~1214~b \citep{charbonneau2009}. This sample is identical to the one explored in \citet{breza25}. The zero-albedo {\teq} varies from $\sim$280~K to $\sim$1250~K within this sample. For TOI-824~b, and TOI-1231~b, we assume a $100\times$solar metallicity due to lack of strong metallicity constraints. Maintaining consistency with recent {\it JWST} observations, we assume different metallicities for the other sub-Neptunes: K2-18~b (100$\times$solar, \citet{hu2025}), TOI-270~d (200$\times$solar, \citet{benneke2024}), TOI-421~b (0.1$\times$solar, \citet{davenport25}), TOI-836~c (200$\times$solar, \citet{wallack24}), GJ~9827~d (300$\times$solar, \citet{piaulet2024}), and GJ~1214~b (300$\times$solar, \citet{kempton23,gao23}).

\subsection{Impact of clouds on the atmosphere-mantle interface}\label{sec:interface}

We couple our cloudy and clear atmosphere models with the  \texttt{SMILE} interior structure model to investigate how these deep atmospheric clouds alter the pressure and temperature conditions at the atmosphere-mantle boundary. Our interior structure model considers a miscible mixture of \ce{H2}, \ce{He}, and \ce{H2O}, an assumption that has been examined in the literature \cite[e.g.,][]{piaulet2025}. We investigate the impact of clouds on three physical aspects of these boundaries: 1) the phase of the interface, 2) the pressure-temperature conditions at the interface, and 3) the chemical behavior at the interface.

\subsubsection{The phase of the atmosphere-mantle interface}

\begin{figure*}
    \centering
    \includegraphics[width=1\linewidth]{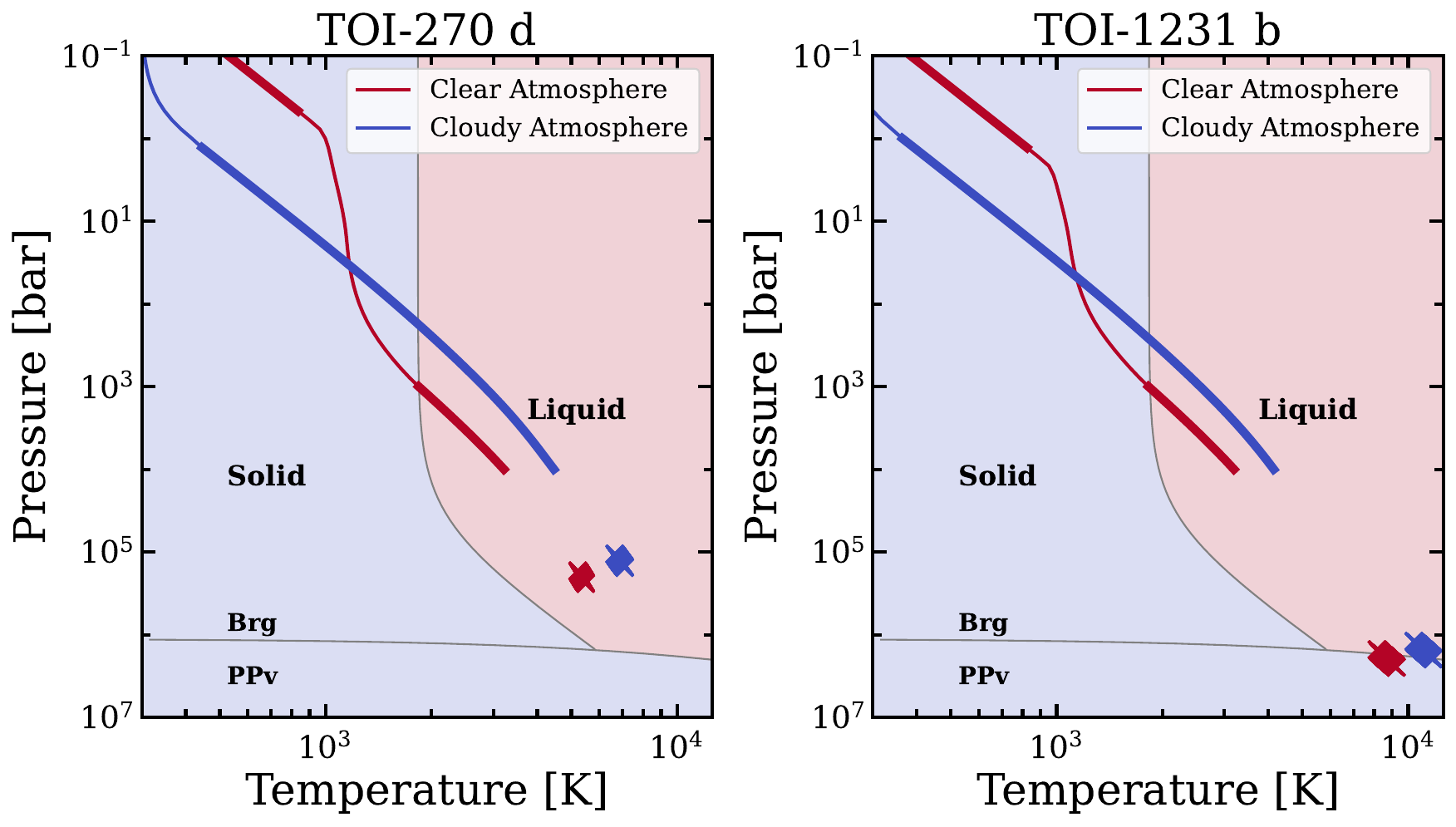}
    \caption{Two panels show the {\tp} profiles for the clear atmosphere (red line) and cloudy atmosphere (blue line) computed for TOI-270~d and TOI-1231~b. The thicker lines depict the convective regions of the atmosphere. The locus of blue and red crosses show the $P$ and $T$ locations of the atmosphere-mantle interface computed by coupling our atmosphere model with the interior structure model for the cloudy and clear models, respectively. The $P-T$ location and the {\tp} profiles are overlaid on the phase--diagram for silicates, depicting its solid and molten phases as blue and red shaded regions, respectively. The bridgmanite–postperovskite (Brg--PPv) boundaries are also shown. The atmosphere-mantle interface are a locus of points instead of a single point in the $P-T$ space because of its sensitivity to the envelope mass fraction $f_{\rm envelope}$. All cloudy models have been computed assuming $f_{\rm sed}=0.5$. The interface conditions for the values of $f_{\rm envelope}$ that are consistent with the measured planet mass and radius are shown. The interfaces of the other 6 planets in our sample along with a zoomed-in view of the interface conditions for TOI-1231~b are shown in Appendix \S\ref{sec:atmosmantleappen}.}
    \label{fig:planets}
\end{figure*}

%We investigate the impact of clouds on the phase of the atmosphere-mantle interface for 8 temperate to hot sub-Neptunes-- K2-18~b \citep{montet2015}, TOI-270~d \citep{Gunther19}, TOI-421~b \citep{carleo2020}, TOI-824~b \citep{burt2020}, TOI-836~c \citep{hawthorn2023}, TOI-1231~b \citep{Burt2021}, GJ~9827~d \citep{Niraula2017}, and GJ~1214~b \citep{charbonneau2009}. We explore the same sample of sub-Neptunes as \citet{breza25}. The zero-albedo {\teq} varies from $\sim$280~K to $\sim$1250~K within this sample. For K2-18~b, TOI-824~b, and TOI-1231~b, we assume a $100\times$solar metallicity due to lack of strong metallicity constraints. Maintaining consistency with recent {\it JWST} observations, we assume different metallicities for the other sub-Neptunes: TOI-270~d (200$\times$solar, \citet{benneke2024}), TOI-421~b (0.1$\times$solar, \citet{davenport25}), TOI-836~c (200$\times$solar, \citet{wallack24}), GJ~9827~d (300$\times$solar, \citet{piaulet2024}), and GJ~1214~b (300$\times$solar, \citet{kempton23,gao23}).   

Figure~\ref{fig:planets} shows {\tp} profiles of cloudy and clear atmospheres and the location of the atmosphere-mantle interface in the pressure-temperature space overlaid on the phase diagram for silicate. The cluster of these interfaces for TOI-270~d and TOI-1231~b are shown in the two panels of Figure~\ref{fig:planets}. The red and blue points mark the interfaces for the clear and cloudy models of each planet, respectively. These are shown as a cluster of interfaces instead of a single $T$ and $P$ value due to varying envelope mass fractions ($f_{\rm envelope}$) in our interior structure model. The interfaces for the other 6 planets in our sample are shown in Appendix \ref{sec:atmosmantleappen} along with a zoomed-in view of the interfaces of TOI-1231~b.

Figure~\ref{fig:planets} shows that the atmosphere-mantle interface are expected to be in the molten phase for TOI-270~d in both the cloudy and clear cases. However, for TOI-1231~b, the cloudy models prefer a molten interface whereas the clear models are in the solid phase. This is because TOI-1231 b’s measured mass and radius are consistent with a relatively high $f_{\rm envelope}$, leading to interface pressures and temperatures high enough that the clear and cloudy models straddle the boundary between the molten and solid phases. Figure~\ref{fig:planets_appen} in Appendix~\ref{sec:atmosmantleappen} show that the phase of the interfaces remain molten for K2-18~b, TOI-421~b, TOI-824~b, TOI-836~c, and GJ~9827~d in both the cloudy and clear cases. The atmosphere-mantle interface is in the molten phase-space for the cloudy models for GJ~1214~b, whereas they are near the solid state phase in the clear models.  Like for TOI-1231~b, this is again due to the high $f_{\rm envelope}$ values consistent with GJ~1214~b's measured mass and radius. Our results suggest that cloud-driven changes in the $T-P$ conditions can have a strong impact on the phase of the interface for sub-Neptunes, especially those with high values of allowed $f_{\rm envelope}$. %The metal-poor atmosphere assumed for TOI-421~b causes the cloudy and clear interfaces to overlap with one another primarily because lower metal enrichment leads to much lower cloud mass mixing ratios (e.g., Figure~\ref{fig:metal}) and a relatively small impact on the $T(P)$ profile. 

\subsubsection{Pressure--temperature conditions at the interface}

\begin{figure*}
    \centering
    \includegraphics[width=1\linewidth]{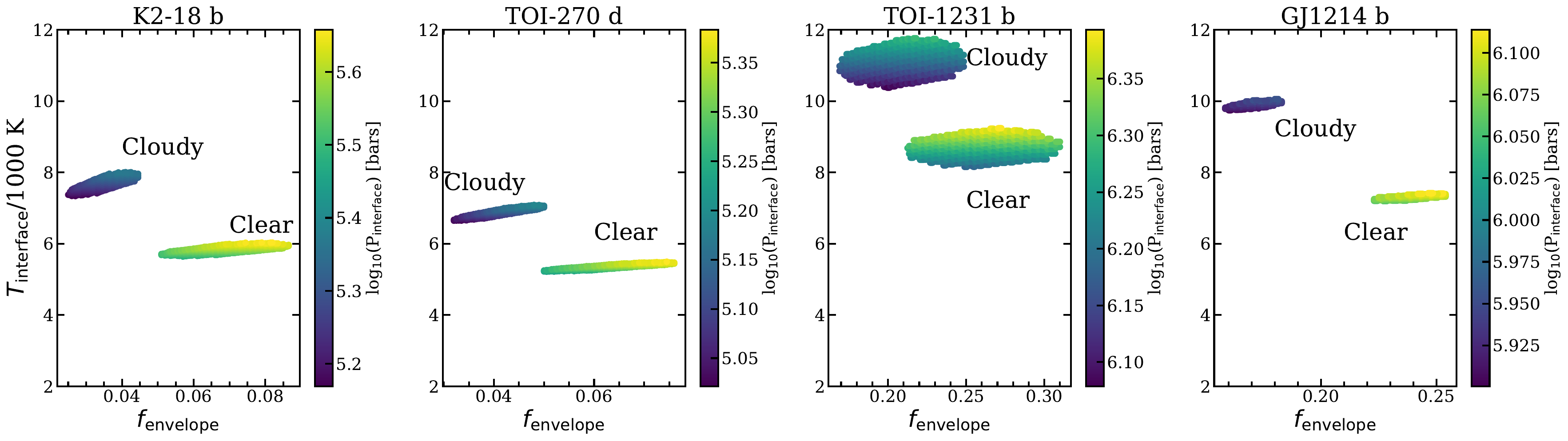}
    \caption{Each panel shows the temperature of the atmosphere-mantle interface as a function of the envelope mass fraction ($f_{\rm envelope}$) for the cloudy and clear models shown in Figure~\ref{fig:planets}. The color of each point depicts the pressure of the atmosphere-mantle interface. The locus of interfaces in the $T-f_{\rm envelope}$ space for the cloudy and clear models are marked. The dependence of the interface temperature and pressure on $f_{\rm envelope}$ for the other four sub-Neptunes in our sample are shown in Appendix~\ref{sec:tpappen}.}
    \label{fig:fenv}
\end{figure*}

For all sub-Neptunes in our sample except TOI-421~b, the atmosphere-mantle interface is pushed to substantially different $T$ and $P$ conditions in the cloudy planet models relative to the clear models. The cloudy atmospheres heat up the envelope to varying levels and move the atmosphere--mantle interface to hotter temperatures and lower pressures, with temperature differences of $\ge{1400}$~K between the cloudy and clear models. Figure~\ref{fig:fenv} shows the interface temperature derived from cloudy and clear atmosphere models as a function of the envelope mass fraction ($f_{\rm envelope}$) for K2-18~b, TOI-270~d, GJ-1214~b, and TOI-1231~b. For K2-18~b, the cloudy model has an interface temperature of about ($\sim$) 7600~K; that is $\sim$1843~K hotter than the clear model ($\sim$5800~K). The interface temperature in the cloudy models show a stronger dependence on $f_{\rm envelope}$ relative to the clear models. For TOI-270~d, the interface temperature in the cloudy model is hotter by $\sim{1517}$~K than the clear model. The interfaces for GJ~1214~b and TOI-1231~b are heated by $\sim{2588}$~K and $\sim{2393}$~K relative to the clear models, respectively. We show the interface conditions for the other four sub-Neptunes in Appendix~\ref{sec:tpappen}. TOI-824~b, TOI-836~c, and GJ 9827~d also show $\ge{1400}$~K cloud-driven temperature increase at their atmosphere-mantle interface among these four planets. The interface for TOI-421~b does not show any significant cloud-driven change in $T-P$ conditions owing to its low atmospheric metal-enrichment (Appendix~\ref{sec:tpappen}). This shows that the temperatures at the atmosphere-mantle interface can be significantly increased due to clouds, particularly in metal-enriched sub-Neptunes.

The pressure conditions at the interface are shown as colors of markers in each panel of Figure~\ref{fig:fenv}. Cloudy models have atmosphere-mantle interfaces at shallower pressures than clear atmosphere models. The difference in interface pressure ranges between $\Delta{\log_{10}{P}}\sim$0.29 (in bar) (K2-18~b) to $\Delta{\log_{10}{P}}\sim$0.06 (in bar) (TOI-824~b), except for TOI-421~b for which there is no difference in interface pressure between cloudy and clear models. The interface pressure also shows a weak dependence on $f_{\rm envelope}$, where it generally tends to increase with higher $f_{\rm envelope}$. However, the extent of this increase varies between planets in our sample. For example, the interface pressure in the clear models of K2-18~b increases by about 42\% when $f_{\rm envelope}$ increases from 0.052 to 0.083. The interface pressure in the cloudy models show about 46\% increase when $f_{\rm envelope}$ increases from 0.026 to 0.044. On the other hand, the increase in interface pressure for TOI-1231~b is about 16\% both in the cloudy and clear models within the extent of allowed $f_{\rm envelope}$ values. GJ~1214~b models show about 9\% increase in interface pressure with increasing $f_{\rm envelope}$ for both cloudy and clear cases. Table~\ref{tab:interface_properties} summarizes the average interface properties across the 8 sub-Neptunes in our representative sample.

\begin{table*}
\centering

%\scriptsize
%\setlength{\tabcolsep}{3pt}
\renewcommand{\arraystretch}{1.2}
\begin{tabular}{c|c|ccc|ccc|cc}
\hline
\hline
Planet 
& M/H 
& \multicolumn{3}{c|}{$T_{\rm interface}$ [K]}
& \multicolumn{3}{c|}{$\log{_{10}}(P_{\rm interface})$ [bar]}
& \multicolumn{2}{c}{Phase} \\
\cline{3-5}
\cline{6-8}
\cline{9-10}
& ($\times$ Solar)
& Clear & Cloudy & $\Delta T$
& Clear & Cloudy & $\Delta \log{_{10}}(P)$
& Clear & Cloudy \\
\hline
K2-18 b    & 100  & 5836  & 7679 & 1843 & 5.56 & 5.27  & -0.29 & molten  & molten \\
TOI-270 d  & 200 & 5353  & 6870 & 1517 & 5.30 & 5.10 & -0.19  & molten & molten \\
TOI-421 b  & 0.1 & 3064 & 3207 & 142 & 4.43 & 4.43 & 0 & molten  & molten \\
TOI-824 b  & 100 & 6804 & 8824 & 2019 & 5.59 & 5.52 & -0.06 & molten & molten  \\
TOI-836 c  & 200 & 6205 & 8552 & 2347 & 5.72 & 5.56 & -0.16 & molten & molten \\
TOI-1231 b & 100 & 8692 & 11085 & 2393 & 6.28 & 6.18 & -0.09 & solid  & molten \\
GJ 9827 d  & 300 & 5107 & 6531 & 1423 & 5.02 & 4.82 & -0.19 & molten & molten \\
GJ 1214 b  & 300 & 7302 & 9891 & 2588 & 6.08 & 5.92 & -0.15 & solid/molten & molten \\
\hline
\end{tabular}
\caption{Average interface temperature and pressure computed for the clear and cloudy models for the representative sample of 8 sub-Neptunes. The difference in the average interface temperature and pressure between the cloudy and clear models are also listed. The last two columns show the phase of the atmosphere-interface for each planet. We note that these are average values computed over the range of allowed $f_{\rm envelope}$ values for each planet.  }\label{tab:interface_properties}

\end{table*}

\subsection{Impact on atmosphere-mantle chemical interactions}\label{sec:chemical}

\begin{figure*}
    \centering
    \includegraphics[width=0.9\linewidth]{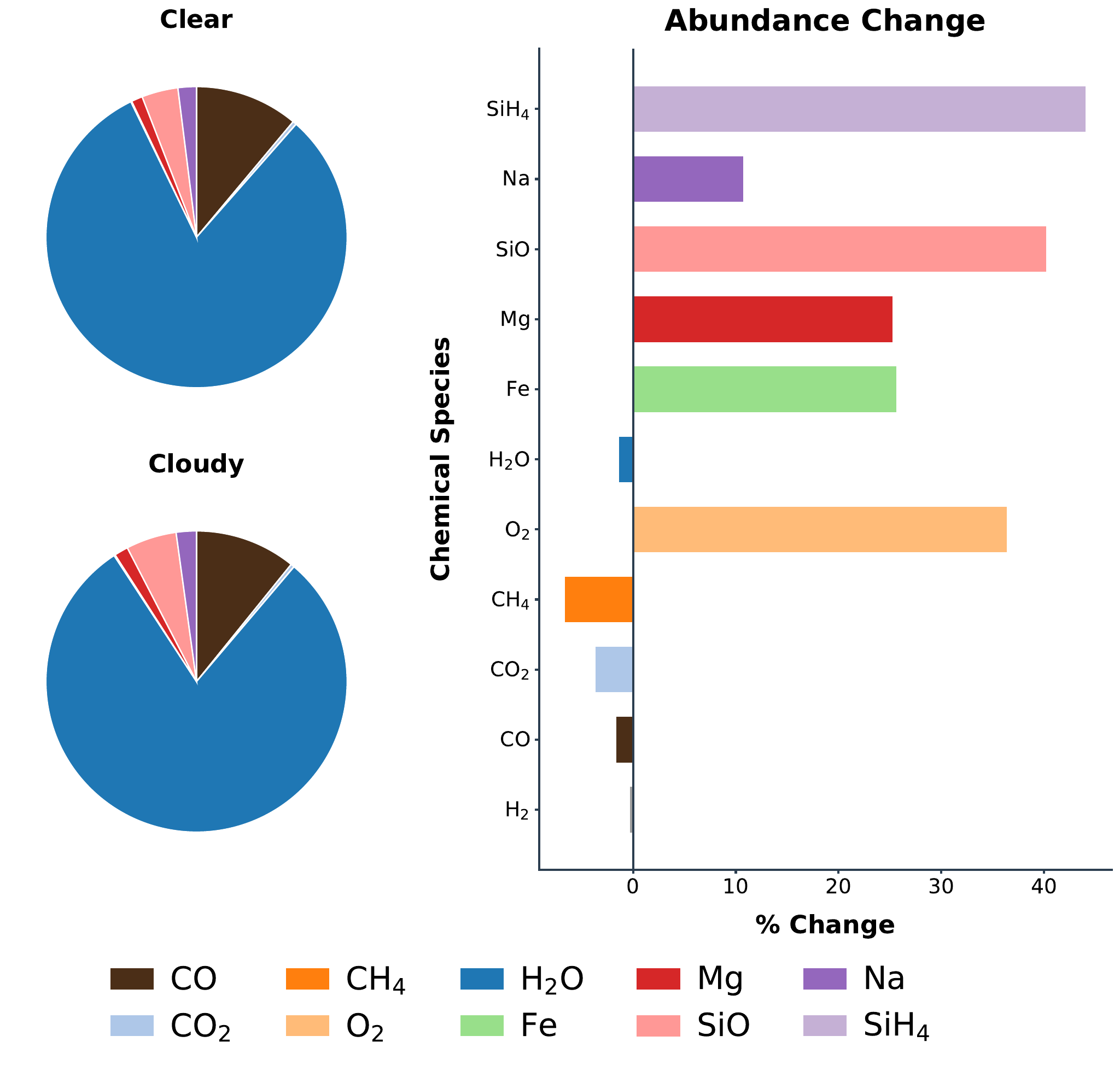}
    \caption{The two pie charts show the fraction of individual molecules at the molten atmosphere-mantle interface for TOI-270~d for the clear (top pie chart) and cloudy (bottom pie chart) models, assuming 50$\times$solar metallicity. The fraction shown in the pie charts have been calculated for all gases except \ce{H2} i.e., fraction of $i$'th gas is defined as $f_i=\dfrac{X_i}{1-{X_{\rm H_2}}}$, where $X_i$ is the abundance of the $i$'th gas at the atmosphere-mantle interface. The pie charts highlights fractions for the dominant non-\ce{H2} gases including \ce{CO}, \ce{H2O}, \ce{Na}, \ce{SiO}, and \ce{Mg}, while gases like \ce{CO2}, \ce{CH4}, \ce{O2}, or \ce{Fe} are present but not visible. The side panel shows the \% change in the abundance of gases in the cloudy model relative to the clear model for each molecule following the same color scheme as the pie charts. A positive value indicates that the gas abundance has increased in the cloudy model relative to the clear model.}
    \label{fig:schlich}
\end{figure*}

We explore how increased temperature at the atmosphere-mantle boundary due to clouds may impact the composition at the base of the atmosphere using the global chemical equilibrium model described in \citet{schlichting2022}. We consider the planet TOI-270~d, a planet whose observed atmospheric composition \citep{benneke24} has been suggested to be consistent with the outcome of magma-atmosphere interactions \citep{Nixon2025}. Hycean conditions for TOI-270~d have not been ruled out \citep{holmberg24} but for our case study we assume the presence of a magma ocean. The global chemical equilibrium model becomes unreliable at temperatures exceeding $\sim$5000~K due to a lack of relevant experimental data; we therefore consider models with a metallicity of 50$\times$ solar for which both the clear and cloudy cases result in temperatures below 5000~K at the atmosphere-mantle boundary. All parameters other than the temperature at the atmosphere-mantle boundary are fixed to those used in \citet{Nixon2025}. For the atmosphere-mantle boundary temperature, we consider two values: 3897 K, the mean boundary temperature recorded for cloud-free atmospheres of TOI-270~d, and 4842~K, the mean temperature for cloudy atmospheres. 

The resulting compositions at the base of the atmosphere are shown as pie charts in Figure~\ref{fig:schlich}. The pie charts show the fraction of different gases excluding \ce{H2}. The fraction of the $i$'th gas is computed by $f_i=\dfrac{X_i}{1-{X_{\rm H_2}}}$. $X_i$ is the abundance of the $i$'th gas. Both pie charts show the most dominant species at the interface other than \ce{H2} is \ce{H2O}. This is followed by \ce{CO}, \ce{SiO}, \ce{Na}, and \ce{Mg} in decreasing order of abundance. Figure~\ref{fig:schlich} also shows the change in the abundance of each gas between the cloudy and the clear model. We find a number of compositional trends with temperature: the abundances of \ce{H2}, \ce{CO}, \ce{CO2}, and \ce{CH4} all decrease in the cloudy model by $\le{10}$\% relative to the clear model. \ce{H2O} abundance shows a decrease of 1.3\% in the cloudy model than the clear model. On the other hand, abundances of \ce{O2}, Fe, Mg, SiO, Na and \ce{SiH4} substantially increase in the cloudy model relative to the clear model. \ce{O2}, \ce{SiH4}, and \ce{SiO} show the maximum increase of 36\%, 44\%, and 40\%, respectively. Both Fe and Mg show 25\% increase and Na increases by 11\%. Overall, the increased temperature at the atmosphere-mantle boundary leads to the release of more oxygen and refractory species from the interior.

These models highlight that deep atmospheric mineral clouds may alter the composition of a sub-Neptune by changing the thermodynamic properties of the interface between the atmosphere and surface. However, the high boundary temperatures predicted by the atmospheric models largely lie above the range for which experimental data exists. Additional laboratory data at higher temperatures \citep[e.g.,][]{Miozzi2025} are critical in order to better understand the relationship between sub-Neptune interiors and atmospheres. 
%Future work could investigate these effects in a fully coupled manner by iterating between the atmospheric temperature profile and the global chemical equilibrium model to produce converged temperature and chemistry profiles; however, the high boundary temperatures predicted by the atmospheric models largely lie above the range for which experimental data exists. Additional laboratory data at higher temperatures \citep[e.g.,][]{Miozzi2025} are critical in order to better understand the relationship between sub-Neptune interiors and atmospheres.

\section{Discussion}\label{sec:disc}

\subsection{Inhibited convection and moist adiabats}

Mean molecular weight gradients due to condensation of mineral clouds can make convection inefficient in deep H-dominated atmospheres \citep{misener2022,Leconte2017,Guillot1994}. This leads to the formation of radiative layers near the cloud base, allowing steeper temperature gradients than the adiabatic lapse rate to be stable against convection. We do not include this physical effect in our atmospheric or interior structure modeling but such radiative layers will only further increase the temperature at the atmosphere-mantle interface for the cloudy models relative to the clear models. Condensation can release latent heat as well, causing convective parts of atmosphere to follow moist adiabats instead of dry adiabats above the cloud base. While the effects of this latent heat release is almost negligible for near-solar composition gas giant planets \citep{woitke2003}, it can be more pronounced for metal-rich sub-Neptunes \citep{Herbort2022} with more than 1 order of magnitude higher mass mixing ratios of clouds (e.g., Figure~\ref{fig:metal}). Latent heat release may make the temperature gradient less steep than the dry adiabats considered in our work. This will lead to smaller atmosphere-mantle interface temperature differences between the cloudy and clear models than predicted in this work. Future work will be needed to quantify the impact of these two physical processes on our predictions.

\subsection{Impact on spectra}\label{disc:spec}
\begin{figure*}
    \centering
    \includegraphics[width=0.92\linewidth]{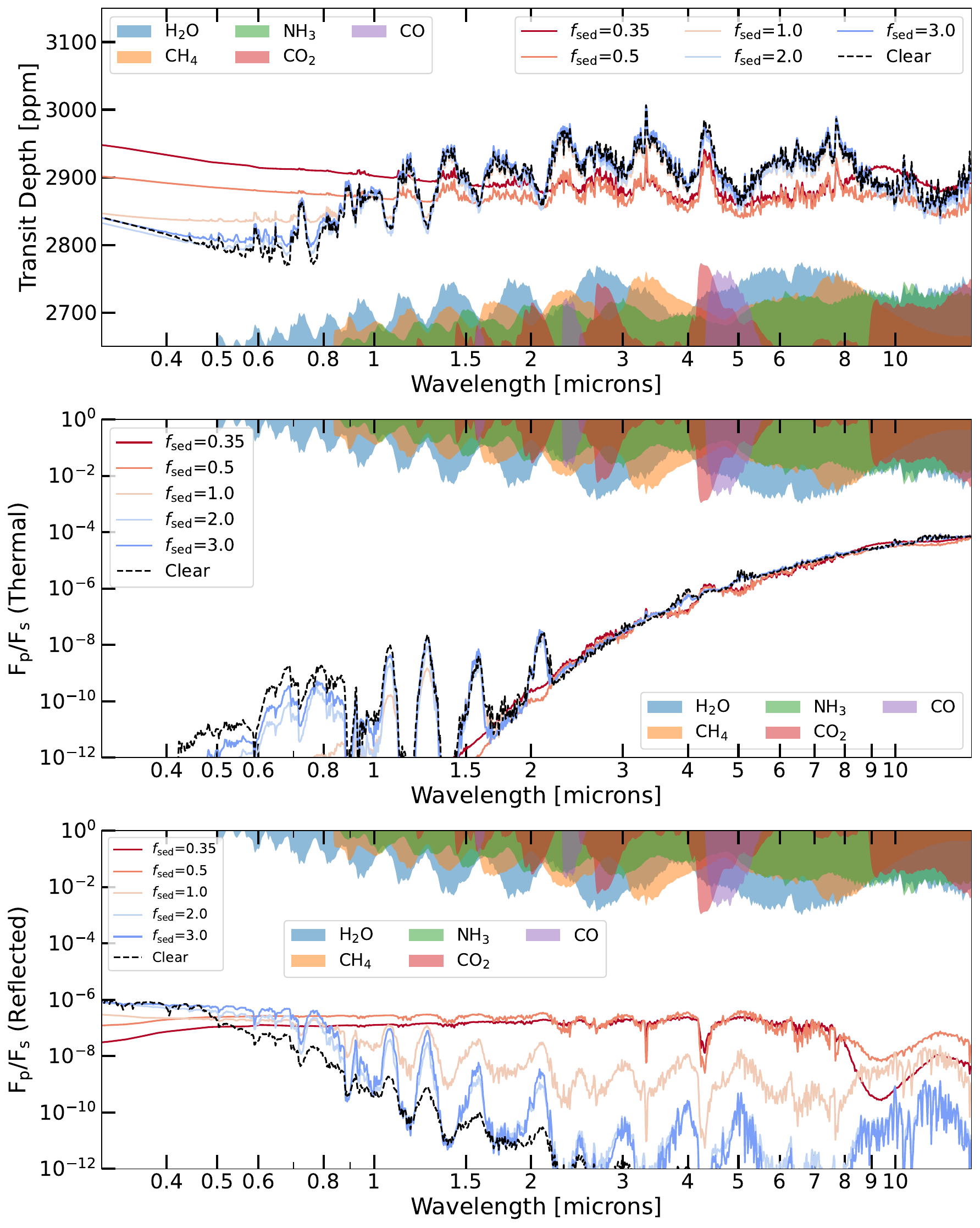}
    \caption{{\it Top panel}: The transmission spectra for the clear and cloudy models for TOI-270~d shown in Figure~\ref{fig:fsed}. These models assume a metallicity of $200\times$Solar. Each colored line shows the spectra for a different $f_{\rm sed}$ value whereas the spectra for the clear atmosphere is shown with the black dashed line. The y-axis shows the transit depth in parts per million. The opacities of major atmospheric absorbers scaled by their abundance in the clear model are shown as filled regions at the bottom for identification of absorption features in the transmissions spectrum. {\it Middle panel}: Same as the top panel but for the thermal component of the eclipse spectrum for TOI-270~d. The y-axis shows the ratio of the planet and stellar flux. {\it Bottom panel}: Same as the middle panel but for the reflected-light component. The middle and bottom panels also show the abundance-scaled opacities at the top in an inverted scale for identification of features.}
    \label{fig:spec}
\end{figure*}
While we have focused on the profound impact of clouds on the physical conditions at the atmosphere-mantle interface, they also shape the spectra of these planets \citep[e.g.,][]{lee2025}. Figure~\ref{fig:spec} shows the impact of clouds on the transmission, emission, and reflected-light spectra of TOI-270~d for the clear atmosphere models and cloudy models with different $f_{\rm sed}$. We note that our goal here is not necessarily to fit the observed transmission spectrum of TOI-270~d \citep[e.g.,][]{benneke2024}, but rather to explore how these clouds modify the spectrum across different observing geometries. A lower $f_{\rm sed}$ causes more muting in the transmission spectrum as cloud particles are lofted to smaller pressures. Lower $f_{\rm sed}$ values also change the chemical composition of gases like {\meth}, {\co}, and {\amon} in the deeper atmosphere due to their heating effect. A hotter deep atmosphere due to lower $f_{\rm sed}$ causes a lower quenched {\meth} abundance. As a result, a lower $f_{\rm sed}$ value causes smaller as well as more muted {\meth} feature in the L-band in Figure~\ref{fig:spec}. The lowest $f_{\rm sed}$ case also shows the silicate absorption feature between 8-10~$\mu$m, detecting which with {\it JWST} will further test the presence of these lofted cloud droplets in the atmospheres of temperate sub-Neptunes like TOI-270~d. 

Middle panel of Figure~\ref{fig:spec} shows lower $f_{\rm sed}$ clouds ($f_{\rm sed}\le$0.5) suppress the thermal flux from TOI-270~d for wavelengths below $\sim{2}$~$\mu$m. The short-wavelength peaks of the emission spectrum disappear for these very cloudy atmospheres, making them emit more like blackbodies. The cloudier atmospheres are brighter than the clear atmospheres at certain long wavelength bands like $\sim{2.1}$-2.4~$\mu$m and $\sim{4.1}$-4.2~$\mu$m. Because transmission spectroscopy alone can be susceptible to the aerosol–mean molecular weight degeneracy, thermal-emission observations of sub-Neptunes at shorter wavelengths, when analyzed together with their transmission spectroscopy, can prove highly complementary for testing the presence of these predicted clouds and constraining their atmospheric properties.

Reflected-light spectra (bottom panel) show very large differences between these cases. The reflected light signal from the clear atmosphere drops from 1~ppm at wavelengths shorter than $\sim{0.5}~{\mu}$m to below $10^{-4}$~ppm for wavelengths longer than $\sim{1}$~${\mu}$m. But the reflected light from the lowest $f_{\rm sed}$ model ($f_{\rm sed}=0.35$) shows $\sim{0.1}$~ppm level signals between 0.6 to 8~$\mu$m with some molecular absorption bands. This suggests that eclipse spectroscopy of these sub-Neptunes, probing both their reflected and thermal light, can prove diagnostic of these deep clouds with upcoming telescopes like the {\it Extremely Large Telescope}, {\it Nancy Grace Roman Telescope}, and the {\it Habitable Worlds Observatory}.

\subsection{Patchiness of clouds}
We do not consider spatially inhomogeneous clouds in our atmosphere models. However, recent {\it JWST} observations have provided a large volume of evidence that aerosol coverage for hot giant exoplanets can be patchy. This evidence has come in the form of large aerosol-driven limb  asymmetries in transmission \citep[e.g.,][]{Mukherjee2025,Fu2025} and phase-curve measurements \citep[e.g.,][]{coulombe2025}. These spatial variations are likely driven by temperature difference-driven diurnal cloud cycles \citep{Mukherjee2025}. But 3D models show that the temperature structure across the globe tend to become more homogeneous for colder exoplanets due to very long radiative-cooling timescales of their atmospheres \citep{roth24}. This suggests that the temperature-driven diurnal cloud cycles may be weaker in colder sub-Neptunes. 

Our models show that clouds cause the radiative-convective boundary to move to smaller pressures relative to clear models. 3D and 1.5D models show that the day and night sides of exoplanets should be thermally homogeneous deeper than the radiative--convective boundary \citep{lines19,gandhi2020}. This suggests that these deep mineral clouds in sub-Neptunes are likely to be much more homogeneous across their day and night sides than in hot Jupiters, where they are formed near relatively smaller pressures. Therefore, our use of a homogeneous cloud-cover is justified in the deep atmosphere, but how these deep mineral clouds affect the circulation patterns in sub-Neptunes need to be further explored with 3D models \citep[e.g.,][]{charnay21}.

\subsection{Implications for thermal evolution}

While several physical processes with potentially large impact on the radius evolution of sub-Neptunes have been identified \citep[e.g.,][]{gupta2019,tang25,lopez14,vazan2018,steinmeyer2026}, the thermal profile of the deep atmosphere remains a key factor that regulates their cooling rate and radius contraction over time. The clouds explored in this study make the atmospheres of these sub-Neptunes more opaque, likely diminishing their efficiency of losing their intrinsic heat. As a result, these clouds are expected to slow down the thermal evolution and radius contraction of these low mass planets.  The adiabatic temperature of the atmosphere at 10~bars or $T_{10}$ has been used in the literature to compute the thermal evolution of irradiated or isolated substellar objects over time \citep[e.g.,][]{Mukherjee2026,diamondback,saumonmarley08}. For the TOI-270~d models shown in Figure~\ref{fig:cloudvsclear}, the adiabatic $T_{10}$ of the cloudy model is hotter by 330 K relative to the clear model. Using clear-atmosphere boundary conditions in thermal evolution models would therefore tend to overestimate the cooling rate of cloudy sub-Neptunes and underestimate the duration over which their atmosphere–mantle interfaces remain molten.  %Clouds might also cause the atmosphere-mantle interface to remain molten for a longer timescale than estimated from evolutionary models without clouds. %Thermal evolution models that are self-consistently coupled to these cloudy atmospheres need to be computed in the future to fully explore the implications of these mineral clouds on sub-Neptune evolution.

\subsection{Bias on intrinsic temperature {\tint}}

Another related implication of the cloud-driven deep atmospheric heating is a potential bias on the measured {\tint} of sub-Neptunes. Quenching of gases like {\meth}, {\co}, and {\amon} has been proposed as a way to measure the intrinsic heat of planets \citep[e.g.,][]{fortney20,ohno23,mukherjee25}. This technique has been used to estimate the intrinsic heat flux of a few giant exoplanets, Neptune-like exoplanets, and sub-Neptunes with {\it JWST} observations \citep{welbanks24,sing24,barat2025,beatty2024}. A key aspect of this technique is that it probes the temperature and pressure conditions of the atmosphere near the quench pressures of gases like {\meth}, {\co}, {\cotwo}, and {\amon}, which is driven by atmospheric vertical mixing. The interpretation of this $T-P$ condition as a measure of the interior heat flux characterized by {\tint} is more complex. Figure~\ref{fig:cloudvsclear} shows clouds might cause the deep atmosphere to be much warmer than clear models, even if both models have the same intrinsic heat flux or {\tint}. This suggests that if Neptune-like or sub-Neptune planets like WASP-107~b, V1298~Tau~b, GJ3470~b, or GJ436~b have such deep atmosphere clouds and we interpret their measured abundances with clear atmosphere models (e.g., RCPE models), we will tend to overestimate the {\tint} of their interiors. This is because a clear atmosphere model will require a higher {\tint} value to match the $T-P$ conditions at the {\meth} quench point to model its observed abundance. This effect could at least partially cause the measured {\tint} of several exoplanets to be much higher than expected from their thermal evolution models \citep{yu2026}. Alternatively, the {\tint} obtained from clear self-consistent atmosphere models can be considered an upper limit.

\subsection{Limitations of modeling the deep interior}

Several of our conclusions regarding the phase of silicates at the atmosphere-mantle boundary, as well as the chemical reactions taking place at that boundary, rely on our relatively limited understanding of material properties at very high pressures and temperatures. For example, this work  incorporates MgSiO$_3$ phase transitions from \citet{Ono2005} and \citet{Belonoshko2005} which have been employed in several internal structure models \citep{Huang2022,breza25}. However, other works have predicted different melting behavior of silicates at high pressure, suggesting that the mantle could remain molten at higher pressures \citep{Fei2021}. Additional experimental work to understand the material properties of planetary components at high pressures is essential in order to more accurately infer sub-Neptune structures.

Our interior-atmosphere exchange models assume a ``reactive'' core in which iron participates in the equilibrium chemistry, as opposed to an ``unreactive'' core which is isolated from interactions with the mantle and atmosphere. \citet{schlichting2022} compared the resulting atmospheric composition from both cases and found some differences, including that CO abundances tend to exceed H$_2$O abundances in the unreactive case, with this effect reversed in the reactive case. However, more recent work by \citet{Young2025} suggests that the cores and mantles of sub-Neptunes are likely to to be fully miscible, suggesting that the iron component of the planet is likely to participate in the equilibrium chemistry. For this reason, we adopted the reactive core throughout this work.

\section{Conclusions}\label{sec:conc}
% We explored the impact of mineral clouds on the atmosphere-mantle interface of sub-Neptune exoplanets. We used a self-consistently cloudy climate model and a interior structure model to quantify how mineral clouds in deep atmospheres of sub-Neptunes might alter the phase, temperatures, pressures, and chemical nature of the atmosphere-mantle boundary. Our key findings are:

 Our work reveals several key findings about the impact of clouds on the atmosphere-mantle boundary of sub-Neptunes. Using a self-consistent cloudy climate model coupled to an interior-structure model and a magma-atmosphere chemistry framework, we find:

\begin{enumerate}
    \item {\bf Cloud-driven heating and cooling}: We find that abundant mineral cloud formation with base pressures between $\sim$100-1000 bars might heat up the deep atmospheres of sub-Neptunes like TOI-270~d by $\ge{1000}$~K at 10$^4$~bar. These clouds, at the same time, can cool down the higher altitudes of the atmosphere by $\sim$600~K near $\sim$1~bar. Such optically thick clouds can have a large impact on the convective structure of the atmosphere of these planets, moving the radiative--convective boundary by several orders of magnitude in pressure. %The cooling caused by these clouds at smaller pressures can also lead to condensation of other gases that were otherwise not predicted from clear models.

    \item {\bf Impact of cloud sedimentation efficiency}: The heating effect of clouds is a strong function of the sedimentation parameter $f_{\rm sed}$, where lower $f_{\rm sed}$ values cause larger heating relative to a clear atmosphere. The heating can be as large as $\sim$2000~K at $10^{4}$~bar for $f_{\rm sed}=0.35$ in TOI-270~d. On the other hand, it can be as small as $\sim$250~K for $f_{\rm sed}=3$.

    \item {\bf Role of atmospheric metallicity}: Atmospheric metallicity controls the extent of mineral cloud formation and the heating resulting from their radiative feedback. The heating at $10^{4}$~bar can vary from $\sim{900}$~K to $\sim{1300}$~K if TOI-270~d's metallicity is increased from $10\times$ to $300\times$solar.

    \item {\bf Effect of clouds on the atmosphere--mantle interface}: The phase of the atmosphere-mantle interface remains molten for both the cloudy and clear models of most sub-Neptunes in our sample, except for TOI-1231~b and GJ~1214~b. For these two planets, the cloudy model prefers a molten boundary whereas the clear model are either in or extremely close to the solid phase region owing to high $f_{\rm envelope}$ values required to explain their mass and radius.

    \item {\bf Quantifying cloud-driven heating of the interface}: Clouds substantially raise the temperature at the atmosphere–mantle interface, by about $\Delta T \sim 1400$–$2600$~K relative to clear atmosphere models across various sub-Neptunes in our sample. TOI-421~b is the exception because of its lower metallicity.

    \item {\bf Impact on interface pressure}: The pressure of the atmosphere-mantle boundary shows changes due to clouds too. The cloudy models generally show smaller interface pressures within $\Delta{\log_{10}{P}}\le$0.3 (in bar).

    \item {\bf Effect on interface chemistry}: The cloud-driven change in $P-T$ conditions of the atmosphere-mantle interface causes large changes in the chemistry at this interface relative to clear models. Molecules like \ce{O2}, \ce{SiH4}, and \ce{SiO} increase in abundance by $\ge{36}\%$ in the cloudy model for TOI-270~d. \ce{Fe}, \ce{Na}, and \ce{Mg} are also enhanced but by more moderate amounts. Molecules like \ce{H2}, \ce{H2O}, \ce{CO2}, \ce{CO}, and \ce{CH4} are depleted at the interface of the cloudy models by $\le{10}\%$. 
\end{enumerate}

% We have outlined several implications of these results in terms of understanding the thermal evolution, interior heat fluxes, and spectra of these sub-Neptune exoplanets. The scope of future modeling, observational, and experimental work on this population remains vast. A key priority would be to build a coupled atmosphere--interior model including cloud radiative feedback, which can in turn serve as the boundary condition for thermal evolutionary models of these sub-Neptunes. The full complexity of this planet population can only be uncovered within such increasingly complex modeling frameworks.

 These results argue that deep mineral clouds should be included in climate models and their impact considered as part of the boundary conditions in coupled atmosphere-interior models of sub-Neptunes. The thermal profile of the deep atmosphere, the phase of the atmosphere-mantle interface, and the chemistry of any exchange between the atmosphere and mantle are all sensitive to the presence of clouds. Using clear atmosphere models in evolutionary and retrieval analysis may bias inferences on the intrinsic heat flux, interior structure, and atmospheric composition. Laboratory measurements at the high temperatures predicted by these cloudy models are needed to extend the present analysis to metal-rich sub-Neptunes. Overall, the scope of future modeling, observational, and experimental work on this population remains vast. A key priority would be to build a coupled atmosphere--interior model including cloud radiative feedback, which can in turn serve as the boundary condition for thermal evolutionary models of sub-Neptunes. The full complexity of this planet population can only be uncovered within such increasingly complex modeling frameworks.

%% Please use the acknowledgment and contribution environments. This will 
%% be anonomyized when the "anonymous" style option is used. 
\begin{acknowledgments}
SM and MCN are supported through the 51 Pegasi b fellowship from the Heising-Simons Foundation. ASU's SOL and PHOENIX supercomputers were used for this work. We thank the anonymous reviewer for feedback that helped us to improve the manuscript. 
\end{acknowledgments}

\begin{contribution}
%%This section gives authors the space to recognize author contributions. The text inside this environment is NOT counted towards the total word quanta. At a minimum, manuscripts are expected to include this text:

SM formulated the idea for this research, performed the cloudy and clear climate simulations, contributed to necessary developments of the \texttt{PICASO} climate model, and led the preparation of this manuscript. MCN performed the interior structure modeling and the atmosphere-interior interface chemistry simulations. LW and MRL provided scientific feedback and mentorship throughout the project. JM, NFW, and NEB led several developments in the \texttt{PICASO} climate model needed for this work.

%% But authors are expected to provide more specific details, e.g. 
%%
%%SC was responsible for writing and submitting the manuscript.
%%WWM came up with the initial research concept and edited the manuscript.
%%OTS obtained the funding and edited the manuscript.
%%EBF provided the formal analysis and validation. He also edited the manuscript.
%%GEH Supervised the undergraduates, wrote the software and administers the project github and Zenodo repositories.
%%
%% Authors can use the Contributor Role Taxonomy (CRediT) at
%% for ideas on how write a good statement tailored to their needs.

\end{contribution}

%\facilities{HST(STIS), Swift(XRT and UVOT), AAVSO, CTIO:1.3m, CTIO:1.5m, CXO}

\software{\texttt{PICASO} \citep{Mukherjee22,mang26,batalha19},  
          \texttt{VIRGA} \citep{rooney21,batalha25}, 
          {\it photochem} \citep{wogan2025,wogan24}
          }, \texttt{SMILE} \citep{Nixon2021}, Magma-atmosphere interaction model \citep{schlichting2022}


\appendix

\section{Phase of atmosphere-mantle boundary for other sub-Neptunes}\label{sec:atmosmantleappen}
\begin{figure*}
    \centering
    \includegraphics[width=1\linewidth]{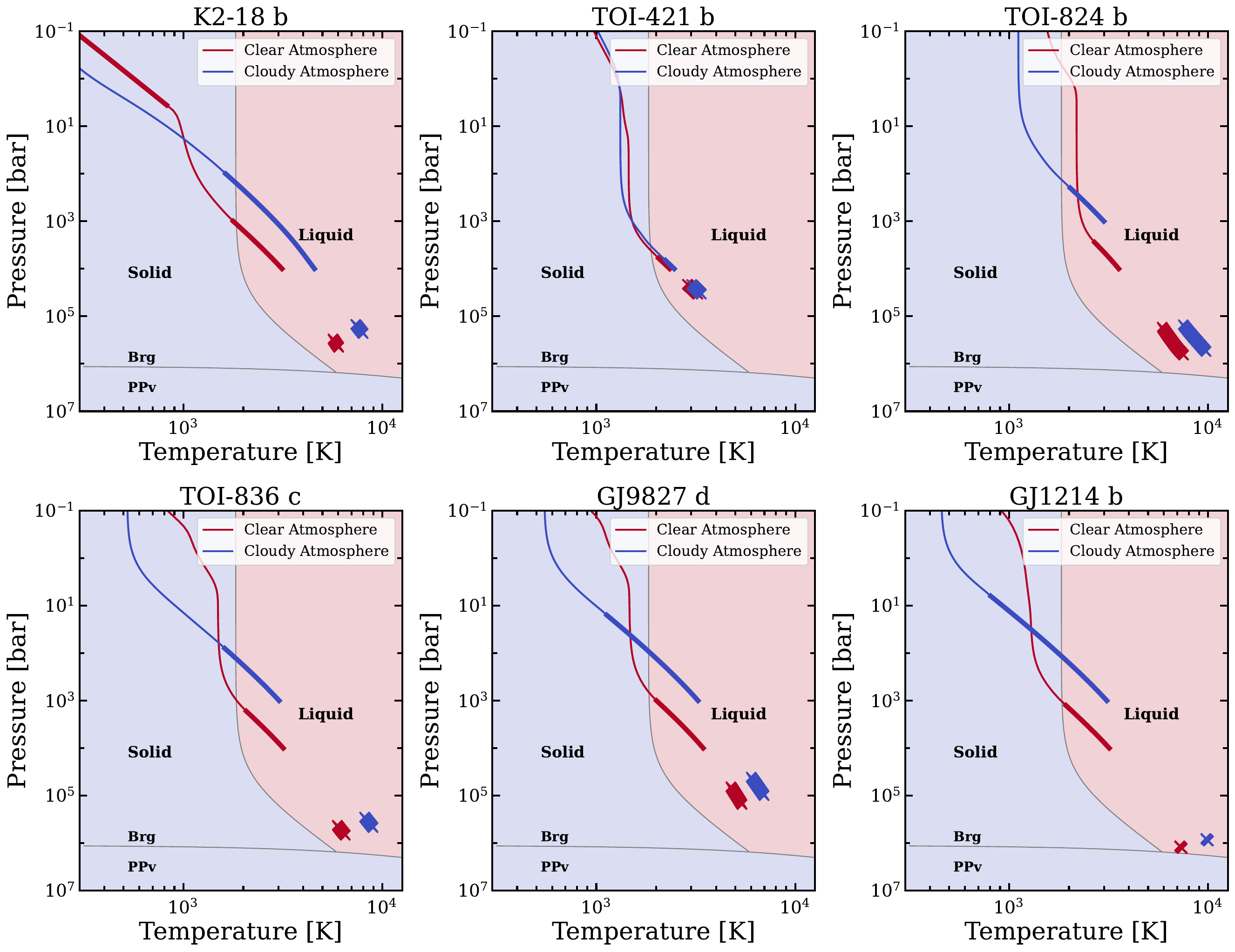}
    \caption{Each panel shows the {\tp} profiles for the clear atmosphere (red line) and cloudy atmosphere (blue line) computed for a sub-Neptune belonging to our sample of 8 sub-Neptunes. The phases of the interfaces for TOI-270~d and TOI-1231~b are shown in Figure~\ref{fig:planets}, while those for the rest of the sub-Neptunes in our sample are shown here. Similar to Figure~\ref{fig:planets}, the thicker lines depict the convective regions of the atmosphere. The locus of blue and red crosses show the $P$ and $T$ locations of the atmosphere-mantle interface computed by coupling our atmosphere model with the interior structure model for the cloudy and clear models, respectively. The $P-T$ location and the {\tp} profiles are overlaid on the phase--diagram for silicates, depicting its solid and molten phases as blue and red shaded regions.}
    \label{fig:planets_appen}
\end{figure*}

\begin{figure}
    \centering
    \includegraphics[width=0.5\linewidth]{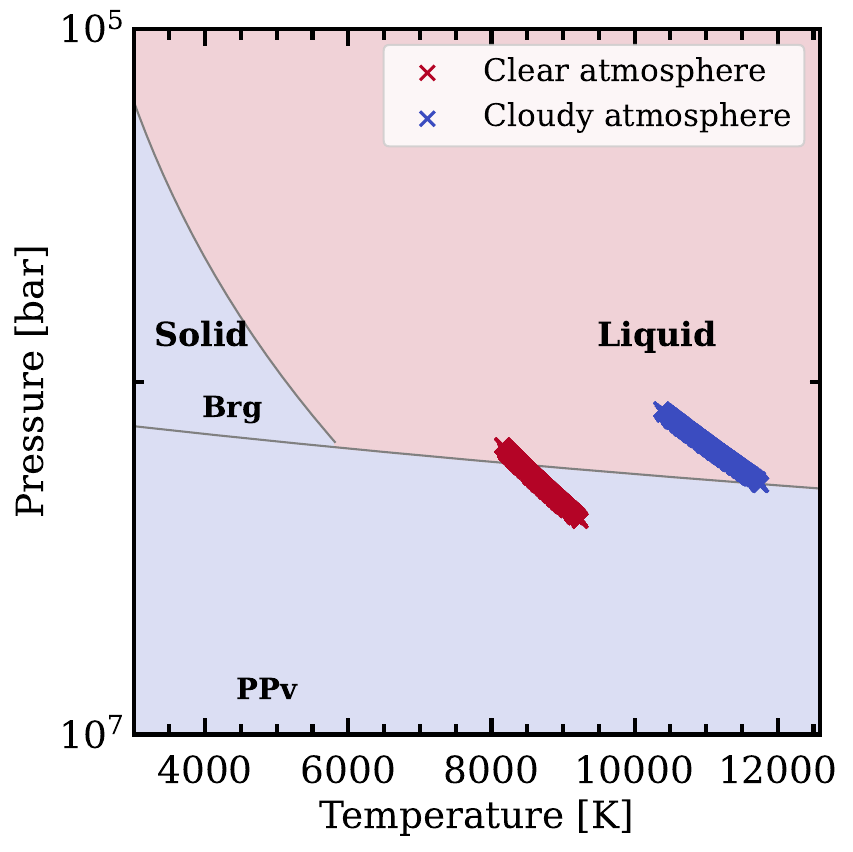}
    \caption{A zoomed-in view of the interface conditions for the cloudy and clear models for TOI-1231~b shown in Figure~\ref{fig:planets}.}
    \label{fig:toi1231b_appen}
\end{figure}

Figure~\ref{fig:planets_appen} show the {\tp} profiles and atmosphere-mantle interfaces of K2-18~b, TOI-421~b, TOI-824~b, TOI-836~c, GJ-9827~d, and GJ-1214~b overlaid on the phase diagram for silicates (similar to Figure~\ref{fig:planets}). The clear models are indicated in red and the cloudy models are indicated in blue. A zoomed-in view of the interface conditions for the cloudy and clear models of TOI-1231~b are shown in Figure~\ref{fig:toi1231b_appen}. The clear models are consistent with a solid atmosphere-mantle interface, whereas the cloudy models are consistent with a molten interface.

\section{$T-P$ conditions of the atmosphere-mantle interface for other sub-Neptunes}\label{sec:tpappen}
\begin{figure*}
    \centering
    \includegraphics[width=1\linewidth]{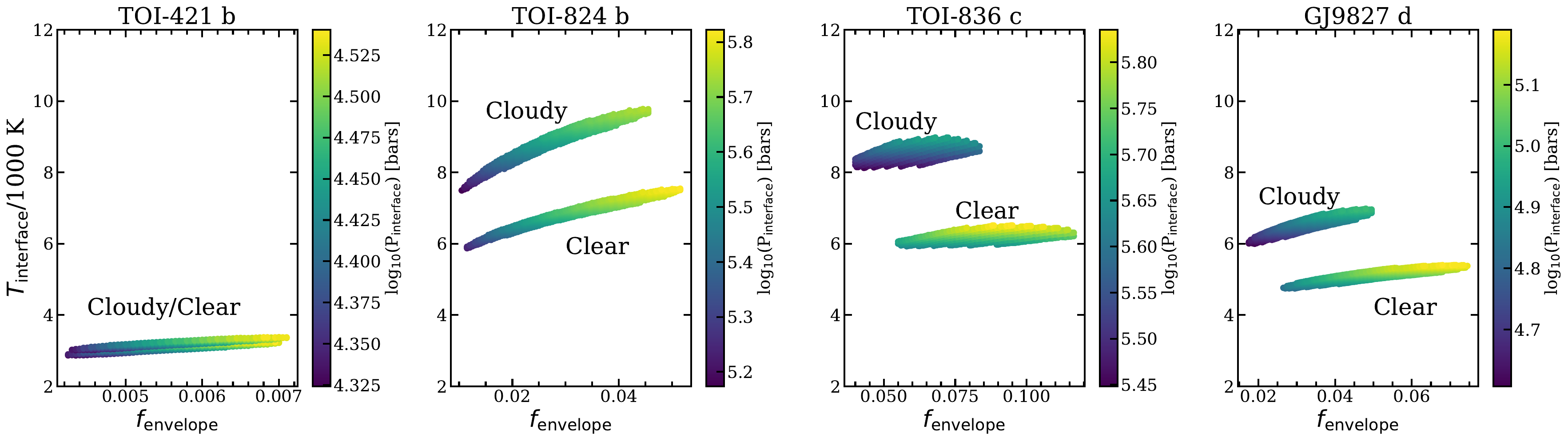}
    \caption{Each panel shows the temperature of the atmosphere-mantle interface as a function of the envelope mass fraction ($f_{\rm envelope}$) for the cloudy and clear models of the sub-Neptunes that are not shown in Figure~\ref{fig:fenv}. The color of each point depicts the pressure of the atmosphere-mantle interface. The locus of interfaces in the $T-f_{\rm envelope}$ space for the cloudy and clear models are marked.}
    \label{fig:fenvappen}
\end{figure*}

Figure~\ref{fig:fenvappen} show the $T$ and $P$ of the atmosphere-mantle interfaces for TOI-421~b, TOI-824~b, TOI-836~c, and GJ~9827~d (similar to Figure~\ref{fig:fenv}). These are the other sub-Neptunes in our sample that have not been shown in Figure~\ref{fig:fenv}. The interfaces for the cloudy and clear models overlap for TOI-421~b due to its low atmospheric metallicity.

\section{Effect of clouds on thermal structure of sub-Neptunes}\label{sec:thermalappen}

\begin{figure*}
    \centering
    \includegraphics[width=1\linewidth]{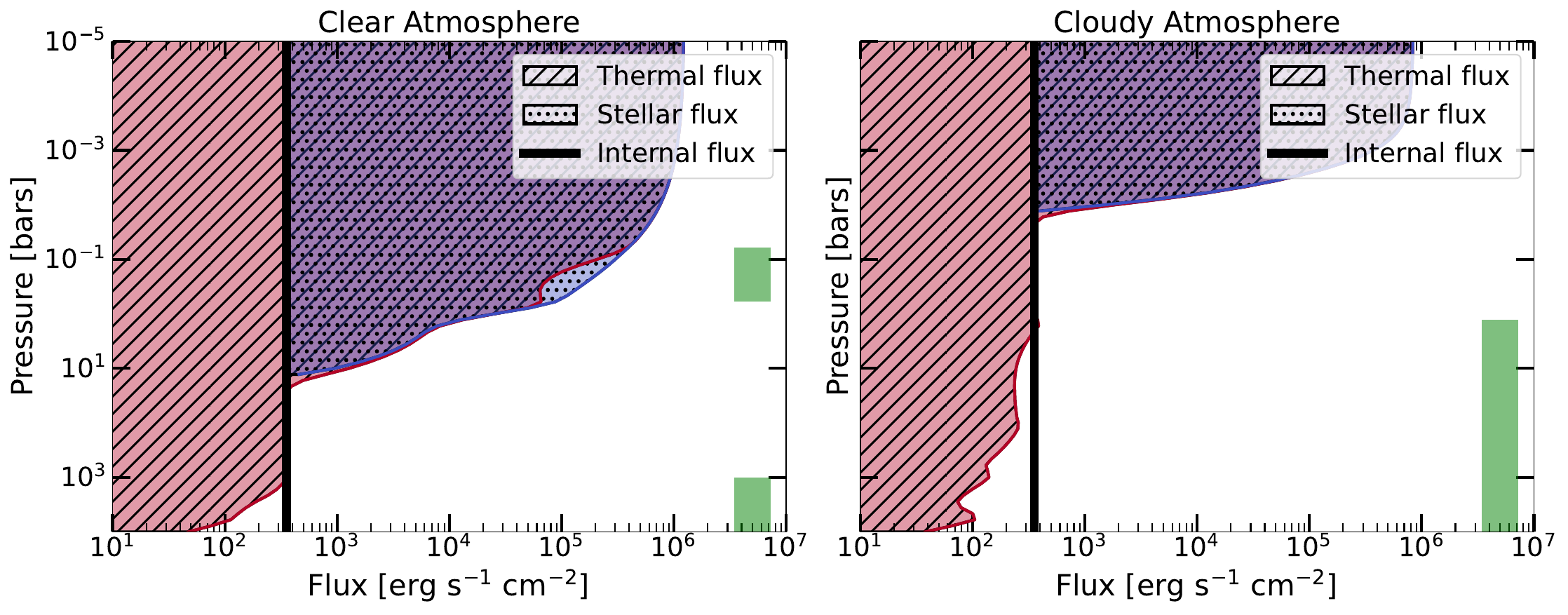}
    \caption{{\it Left panel}: Variation of the net thermal flux (maroon hatched region), net incident stellar flux (dotted blue region), and net intrinsic flux (black line) with pressure for the clear atmosphere model of TOI-270~d shown in Figure~\ref{fig:cloudvsclear}. The convective pressures are shown with the green bars on the right side. {\it Right panel}: Same as the left panel but for the cloudy model. }
    \label{fig:flux}
\end{figure*}

Figure~\ref{fig:flux} shows the impact of clouds on TOI-270~d's climate by comparing the layer-by-layer thermal flux, stellar flux, and intrinsic heat flux between the clear (left panel) and the cloudy (right panel) atmosphere models following \citet{marley15}. The incident stellar flux is fully absorbed near $\sim$10~bar in the clear atmosphere. The planet's thermal emission transfers its intrinsic heat flux between \numrange{10}{1000}~bar. The intrinsic flux is transferred convectively deeper than this pressure in the clear atmosphere. However, the incident stellar flux in the cloudy atmosphere is fully absorbed or reflected by the clouds near a much smaller pressure $\sim$${10^{-2}}$~bar (or higher altitude). This causes the cloudy atmosphere to be colder than the clear atmosphere at pressures between $\sim$$10^{-2}$~bar \& 30~bar. Thermal radiation carries the planet's intrinsic heat flux between \numrange{0.01}{1}~bar, but deeper than $\sim$1~bar convective flux dominates. This extended convective region in the cloudy model is caused by enhanced opacity in the deeper atmosphere due to the clouds. We use wavelength-dependent optical properties of these clouds from the literature: \ce{MgSiO3} from \citet{scott96}, \ce{Fe} from  \citet{palik1991}, \ce{Al2O3} from \citet{koike95,begemann1997}, \ce{Na2S} from \citet{khachai2009,montaner1979}, \ce{KCl} from \citet{querry1987optical}, \ce{MnS} from \citet{montaner1979,huffman1967}, and \ce{ZnS} from \citet{querry1987optical}.

\bibliography{sample701}{}
\bibliographystyle{aasjournalv7}



\end{document}